\documentclass[hyper,letterpaper,12pt]{article}
\usepackage{a4wide}
\usepackage{amsmath}
\usepackage[
      colorlinks=true,
      linkcolor=blue,
      urlcolor=blue,
      filecolor=black,
      citecolor=blue,
            ]{hyperref}
\usepackage{color}
\usepackage{graphicx}
\usepackage{amsfonts}
\usepackage{amssymb}
\usepackage{epstopdf}

\def\beq{\begin{equation}}
\def\eeq{\end{equation}}

\begin{document}

\title{\bf \Large  Entanglement Entropy and Wilson Loop in St\"{u}ckelberg Holographic
Insulator/Superconductor Model}

\author{\large
~Rong-Gen Cai\footnote{E-mail: cairg@itp.ac.cn}~,
~~Song He\footnote{E-mail: hesong@itp.ac.cn}~,
~~Li Li\footnote{E-mail: liliphy@itp.ac.cn}~,
~~Li-Fang Li\footnote{E-mail: lilf@itp.ac.cn}\\
\\
\small State Key Laboratory of Theoretical Physics,\\
\small Institute of Theoretical Physics, Chinese Academy of Sciences,\\
\small P.O. Box 2735, Beijing 100190, China\\}
\date{\small September 05, 2012}
\maketitle

\begin{abstract}
\normalsize We study the behaviors of entanglement entropy and
vacuum expectation value of Wilson loop in the St\"{u}ckelberg
holographic insulator/superconductor model.  This model has rich
phase structures depending on model parameters. Both the
entanglement entropy for a strip geometry and the heavy quark
potential from the Wilson loop show that there exists a
``confinement/deconfinement" phase transition. In addition, we find
that the non-monotonic behavior of the entanglement entropy with
respect to chemical potential is universal in this model. The pseudo
potential from the spatial Wilson loop also has a similar
non-monotonic behavior. It turns out that the entanglement entropy
and Wilson loop are good probes to study the properties of the
holographic superconductor phase transition.
\end{abstract}

\tableofcontents

\section{ Introduction}

The entanglement entropy is expected to be a key quantity to
understand some properties in quantum field theories and in
many-body physics (see, for example,
Refs.~\cite{2006PhRvB..73x5115R,Amico:2007ag}). For a given system,
the entanglement entropy of one subsystem with its complement is
defined as the von Neumann entropy. The entanglement entropy of the
subsystem measures how the subsystem and its complement are
correlated each other. As a von Neumann entropy, the entanglement
entropy is also directly related to the degrees of freedom of the
system. In addition, in quantum many-body physics, the entanglement
entropy is a good quantity to characterize different phases and
associated phase transitions. However, the calculation of
entanglement entropy is very difficult except for the case in $1+1$
dimensions. In the spirit of the AdS/CFT
correspondence~\cite{Maldacena:1997re,Gubser:1998bc,Witten:1998qj,Aharony:1999ti},
a holographic proposal to calculate the entanglement entropy for
strongly coupled field theories has been presented in
Ref.~\cite{Ryu:2006bv} (for reviews see
\cite{Nishioka:2009un,Takayanagi:2012kg}).

On the other hand, while various aspects of holographic
superconductor models have been intensively studied (see, for
example,
Refs.~\cite{Hartnoll:2008vx,Albash:2008eh,Cai:2011tm,Horowitz:2008bn,Brynjolfsson:2009ct,Cai:2009hn,
Horowitz:2009ij,Horowitz:2011dz,Wang:2012yj,Montull:2011im,Bobev:2011rv,Liu:2012hc,Erdmenger:2011tj,
Gubser:2008wv,Hartnoll:2012pp,Erdmenger:2012ik,Montull:2012fy,Barclay:2010up,Cai:2010cv,Siani:2010uw,Cai:2010zm,Wu:2010vr}),
the study of entanglement entropy in the holographic superconducting
phase transition is just initialed.
Refs.~\cite{Albash:2012pd,Cai:2012nm} studied the behaviors of
entanglement entropy for a strip geometry in the holographic s-wave
and p-wave conductor/superconductor models. It turns out that the
entanglement entropy is a good probe to investigate the holographic
phase transition, the entanglement entropy behaves like the thermal
entropy of background black holes, it can indicate not only the
appearance, but also the order of the phase transition.  In a recent
paper~\cite{Cai:2012sk}, the authors investigated the behavior of the
entanglement entropy for a strip geometry in a simple holographic
 insulator/superconductor model at zero temperature. In this
model, the phase transition is a second order one.  It was found that
the entanglement entropy as a function of chemical potential is not
monotonic in the superconducting phase. The entanglement entropy at
first increases and reaches its maximum at a certain chemical
potential and then decreases monotonically as chemical potential
increases.  The non-monotonic behavior of entanglement entropy in
the superconducting phase looks strange. Due to the lack of details
of dual field theory, we did not give a convinced interpretation,
although some possible causes on this behavior were discussed
there~\cite{Cai:2012sk}.



In the holographic insulator/superconductor
model~\cite{Nishioka:2009zj},  the normal insulator phase is
described by a pure AdS soliton solution. As one increases the
chemical potential, the pure AdS soliton background will become
unstable to develop some kind of scalar ``hair". The emergence of
scalar ``hair" induces the symmetry breaking and gives a finite
vacuum expectation value of the dual operator in the field theory
side. The soliton solution with scalar ``hair"  describes a
superconducting phase. The scalar ``hair"  therefore plays the role
of the order parameter in the holographic phase transition.

In this paper we are going to further study the behavior of the
entanglement entropy in the holographic insulator/superconductor
phase transition by generalizing the discussion to the case in the
St\"{u}ckelberg holographic insulator/superconductor
model~\cite{Franco:2009yz}. The aim is two-fold. On the one hand, we
would like to see whether the non-monotonic behavior is universal or
not. On the other hand, the phase structure is rather rich in the
St\"{u}ckelberg holographic insulator/superconductor model. As we
will see shortly, depending on the model parameters, the
superconducting phase transition could be second order or first
order, and further a first order phase transition also occurs in the
superconducting phase.

The St\"{u}ckelberg holographic insulator/superconductor model is
labeled by two parameters. One is  $\beta$ determining the strength
of the back reaction. The other is $\zeta$ determining the form of
interaction between the scalar field and Maxwell field. Both of them
can change the order of the phase transition. As $\zeta$ is
vanishing, an equivalent model has been studied in
Ref.~\cite{Horowitz:2010jq}. It shows that as one increases the
strength of the back reaction, the order of the phase transition is
changed from second order to first order. For the intermediate
strength of the back reaction, although the insulator/superconductor
transition is second order, a new phase transition emerges in
superconducting phase. We fix the strength of the back reaction and
study the effect with the change of $\zeta$. Interestingly, the
resulting phenomenon is qualitatively similar to the above case. The
order of the phase transition is second order for small $\zeta$ and
first order for large $\zeta$. For the intermediate value of
$\zeta$, the phase transition is second order, however, the grand
potential in superconducting phase develops a ``swallow tail"
indicating a new phase transition.

We calculate the entanglement entropy for a strip geometry in this
model, and find that the entanglement entropy can indicate not only
the appearance of phase transition, but also the order of the phase
transition.  Further, no matter the order of the phase transition,
 the entanglement entropy versus chemical potential is always
non-monotonic in the superconducting phase. More precisely, at the
beginning of the transition, the entropy increases and reaches its
maximum at a certain chemical potential and then decreases
monotonically. It indicates that the non-monotonic behavior of
entanglement entropy as a function of chemical potential is
universal in the holographic s-wave insulator/superconductor model.

Apart from the entanglement entropy, there is another nonlocal
quantity, Wilson loop, which can describe some properties of gauge
field theories. In the AdS/CFT correspondence, the vacuum
expectation value (VEV) of Wilson loop  can also be calculated
holographically. In order to give further insights into the
holographic insulator/superconudctor phase transition, we study the
behaviors of temporal Wilson loop and  spatial Wilson loop across
the holographic phase transition. From the VEV of Wilson loops, we
extract the heavy quark potential which describes the interaction
strength  between quark and antiquark.  From the entanglement
entropy and the heavy quark potential, it shows that there exists a
``confinement/deconfinement" phase transition  in the dual field
theory describing the insulator/superconductor model. In addition,
 different from the
phenomenon observed in the entanglement entropy, the (pseudo) heavy
quark potential versus chemical potential in the superconducting
phase will show the monotonic behavior or non-monotonic behavior,
depending on the model parameters.

This paper is organized as follows. In
section~\eqref{sect:stuckelbeg}, we introduce the St\"{u}ckelberg
holographic model. In section~\eqref{sect:smthermo}, the
insulator/superconductor phase transition and its thermodynamics are
investigated in detail. Section~\eqref{sect:smhee} is
 devoted to investigating the behavior
of entanglement entropy in the holographic model. In
section\eqref{sect:wilson}, we study
 the behavior of temporal Wilson loop and spatial Wilson loop with respect to chemical potential
 and distance between quark and antiquark. The conclusion and some discussions are
 included in section \eqref{sect:conclusion}.


\section{St\"{u}ckelberg Insulator/Superconductor Model}
\label{sect:stuckelbeg}

The  St\"{u}ckelberg holographic superconductor model which contains
a real scalar field $\psi$, a real pseudoscalar field $p$ and a
Maxwell gauge field
reads~\cite{Franco:2009yz,Aprile:2009ai,Peng:2011gh}
\begin{equation}\label{action2}
\begin{split}
S &=\int d^5 x\sqrt{-g}[\frac{1}{2\kappa^2}(\mathcal{R}+\frac{12}{L^2})+\frac{1}{\tilde{g}^2}L_{matter}],\\
L_{matter}=-\frac{1}{4}F_{\mu\nu}& F^{\mu \nu}-\nabla_\mu\psi\nabla^\mu\psi-m^2\psi^2-|\mathcal{F}(\psi)|(\nabla_\mu p-A_\mu)(\nabla^\mu p-A^\mu),
\end{split}
\end{equation}
where $\kappa$ is the gravitational constant and $\mathcal{F}$ is a function of $\psi$. In this paper we
consider a simple case $\mathcal{F}(\psi)=\psi^2+\zeta\psi^6$, where
$\zeta$ is a model parameter determining the interaction form and is
assumed to be non-negative to ensure the positivity of the kinetic
term for $p$. For other forms of ${\cal F}$, see \cite
{Franco:2009yz,Aprile:2009ai,Peng:2011gh}. The local $U(1)$ gauge
symmetry in this theory is given by
\begin{equation}\label{gauge}
p\rightarrow p+\theta (x^\mu), \quad A_\mu\rightarrow A_\mu+\nabla_\mu \theta (x^\mu).
\end{equation}
We use this symmetry to set $p=0$ from now on. Here we define a parameter $\beta\equiv\kappa/\tilde{g}$
which measures the strength of the back
reaction of the matter fields on the background geometry.

The equations of motion coming from the above action are
\begin{equation}\label{eom2}
\begin{split}
\mathcal{R}_{\mu\nu}-\frac{\mathcal{R}}{2}g_{\mu\nu}-\frac{6}{L^2}g_{\mu\nu}=\beta^2[F_{\mu\lambda}{{F_\nu}^\lambda}+2\nabla_\mu&\psi\nabla_\nu\psi+2(\psi^2+\zeta\psi^6)A_\mu A_\nu+g_{\mu\nu}L_{matter}],\\
\nabla_\mu\nabla^\mu\psi-m^2\psi&=(\psi+3\zeta\psi^5)A_\mu A^\mu,\\
\nabla_\nu F^{\nu\mu}&=2(\psi^2+\zeta\psi^6)A^\mu.
\end{split}
\end{equation}
Our ansatz for the metric and matter fields are given by
\begin{equation}\label{metric2}
d s^2 =\frac{L^2}{r^2}\frac{d r^2}{g(r)} + r^2(-f(r)d t^2+d x^2+d y^2+g(r)e^{-\chi(r)}d \eta^2),
\end{equation}
\begin{equation}\label{gauge}
\psi=\psi(r),\qquad A=\phi(r) dt,
\end{equation}
where $g(r)$ vanishes at the tip of the soliton $r=r_0$. Further, in order to avoid a conical singularity at the tip $r_0$, $\eta$ should be made with an identification
\begin{equation}
\eta\sim\eta+\Gamma,\qquad \Gamma=\frac{4\pi L e^{\frac{\chi(r_0)}{2}}}{r_0^2 g'(r_0)}\;.
\end{equation}
The independent equations of motion in terms of the above ansatz are
deduced as follows.
\begin{equation}
\label{eomstu}
\begin{split}
&\psi''+(\frac{5}{r}+\frac{f'}{2f}+\frac{g'}{g}-\frac{\chi'}{2})\psi'+\frac{L^2\phi^2}{r^4 fg}(\psi+3\zeta\psi^5)-\frac{L^2 m^2}{r^2 g}\psi=0, \\
&\phi''+(\frac{3}{r}-\frac{f'}{2f}+\frac{g'}{g}-\frac{\chi'}{2})\phi'-\frac{2L^2\phi}{r^2 g}(\psi^2+\zeta\psi^6)=0, \\
&f''+(\frac{2}{r}-\frac{f'}{2f}+\frac{\chi'}{2})f'+(\frac{3\chi'}{r}+4\beta^2\psi'^2)f-\frac{2\beta^2\phi'^2}{r^2}=0,\\
&(\frac{3}{r}-\frac{f'}{2f})g'+(\frac{12}{r^2}+\frac{f'\chi'}{2f}+2\beta^2\psi'^2+\frac{\beta^2\phi'^2}{r^2 f})g+\frac{2L^2\beta^2\phi^2}{r^4 f}(\psi^2+\zeta\psi^6)+\frac{2L^2\beta^2 m^2\psi^2-12}{r^2}=0,\\
&(\frac{3f}{r}+f')\chi'-(\frac{3}{r}+\frac{g'}{g})f'+\frac{4L^2\beta^2\phi^2}{r^4 g}(\psi^2+\zeta\psi^6)
+4\beta^2 f\psi'^2=0.
\end{split}
\end{equation}
In our numerical calculations, we  choose $m^2 L^2=-\frac{15}{4}$
and work in units with $L=1$. However, this analysis can be
generalized to any mass case above the Breitenloher-Freedman bound
$m^2 L^2>-4$. In order to match the asymptotical AdS boundary, the
matter and metric fields near the boundary $r\rightarrow\infty$
should have the form
\begin{equation} \label{boundary2}
\begin{split}
\psi=\frac{\psi_0}{r^{3/2}}+\frac{\psi_1}{r^{5/2}}+\ldots,&\quad \phi=\phi_0-\frac{\phi_2}{r^2}+\ldots, \\
\quad f=1+\frac{f_4}{r^4}+\ldots,\quad
g=1+&\frac{g_4}{r^4}+\ldots,\quad \chi=\frac{\chi_4}{r^4}+\ldots,
\end{split}
\end{equation}
where  $\psi_0$, $\psi_1$, $\phi_0$, $\phi_2$, $f_4$, $g_4$ and
$\chi_4$ are all constants. It is well-known that in
five-dimensional AdS space-time, when $-4<m^2 L^2<-3$, the scalar
field admits two different quantizations related by a Legendre
transformation. $\psi_0$ can either be identified as a source or an
expectation value. In this paper we consider it as source and set
$\psi_0=0$ to accomplish spontaneous symmetry breaking. The
quantities $\phi_0$, $\phi_2$ and $\psi_1$ are related to the
chemical potential $\mu$, charge density $\rho$ and the vacuum
expectation value of the scalar operator $\mathcal{O}$ which has
scaling dimension $\Delta=\frac{5}{2}$ in the dual field theory on
the boundary, i.e., $\mu=\phi_0, \rho=\frac{2\beta^2}{\kappa^2
L}\phi_2$ and
$\langle\mathcal{O}\rangle=\frac{2\beta^2(2\Delta-4)}{\kappa^2
L}\psi_1=\frac{2\beta^2}{\kappa^2 L}\psi_1$.

\section{Thermodynamics and Phase Transition}
\label{sect:smthermo}

In gauge/gravity duality the grand potential (Gibbs free energy)
$\Omega$ of the boundary thermal state is identified with
temperature times the on-shell bulk Euclidean action. Namely $\Omega
=TS_{Euclidean}$. The Euclidean action must include the
Gibbons-Hawking boundary term for a well-defined Dirichlet
variational principle and further a surface counter term for
removing divergence
\begin{equation}
S_{Euclidean}=-\int d^5 x
\sqrt{g}[\frac{1}{2\kappa^2}(\mathcal{R}+\frac{12}{L^2})+\frac{1}{\tilde{g}^2}L_{matter}]
-\frac{1}{\kappa^2}\int_{r\rightarrow\infty} d^4x
\sqrt{h}\mathcal{K}+S_{ct},
\end{equation}
where $h$ is the induced metric on the boundary $r\rightarrow\infty$, and $\mathcal{K}$ is the trace
of the extrinsic curvature. $S_{ct}$ is the counter term given by
\begin{equation}
S_{ct}=\frac{1}{\kappa^2}\int_{r\rightarrow\infty} d^4x
\sqrt{h}(\frac{3}{L}+\beta^2\frac{4-\Delta}{L}\psi^2).
\end{equation}

For our soliton solution (\ref{metric2}), there is no horizon and
associated Hawking temperature and thermal entropy vanish. But for
the Euclidean sector of the solution (\ref{metric2}), one can
introduce an arbitrary inverse temperature $(1/T)$ as the period of
the Euclidean time coordinate.  Due to the fact that the soliton
solution is static, the integration over the Euclidean time in the
Euclidean action  gives an inverse temperature factor $1/T$, which
just cancels the temperature factor in the grand potential and leads
a finite grand potential. Using the on-shell condition and the
expansion of matter and metric functions at infinity
$r\rightarrow\infty$, the grand potential $\Omega$ is found to be
given as
\begin{equation}\label{grand1}
\frac{2L\kappa^2\Omega}{V_3}=g_4,
\end{equation}
with $V_3=\int dx dy d\eta$. Since we have scaled $\Gamma$ to be
$\pi L$, one has $g_4=-1$ in the normal insulator phase.
\begin{figure}[h]
\centering
\includegraphics[scale=0.7]{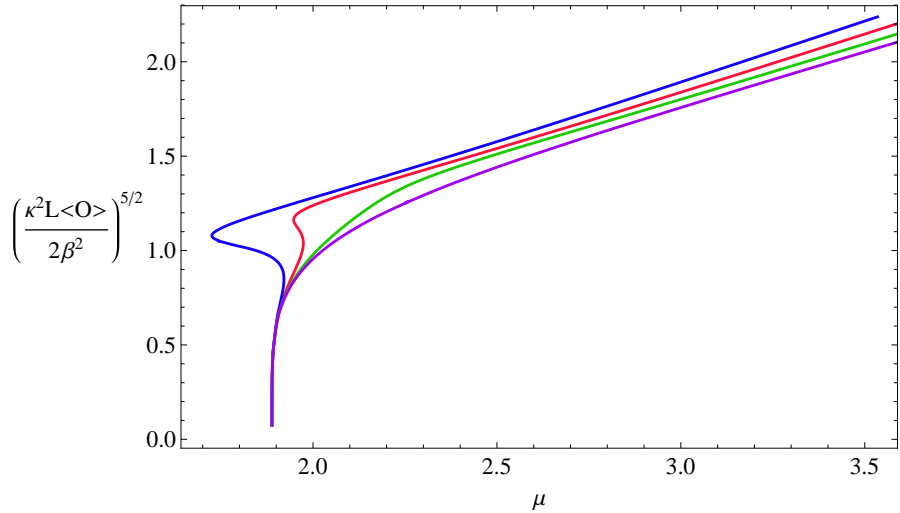}
\caption{\label{smcondensate} The condensate of operator $\mathcal{O}$ versus
chemical potential $\mu$ for $\beta=0.25$. Lines from right to left are for
$\zeta=0$ (purple), $\zeta=0.2$ (green), $\zeta=0.65$ (red) and $\zeta=1.6$ (blue),
respectively. $\Gamma$ is scaled to be $\pi L$.}
\end{figure}

Figure~\eqref{smcondensate} shows the behavior of the condensate
with respect to
 chemical potential for the back reaction parameter $\beta=0.25$\footnote{In our numerical calculations
 of this paper we always
 take $\beta=0.25$ and show rich phase structure by changing the
 parameter $\zeta$.}. When $\zeta$ is small,
 we see from Figure~\eqref{smcondensate} that as the chemical potential $\mu$ exceeds
  a critical value $\mu_c$ the condensate emerges, which implies a superconducting phase appears.
  On the other hand, when $\mu<\mu_c$, the scalar hair vanishes. This is identified as
  the insulator phase, since the system has a mass gap $\sim 1/\Gamma$.
   The insulator/superconductor phase transition here is typically second order and in this case, the
    critical chemical potential $\mu_c$ does not depend on $\zeta$,
    the critical  chemical potential is given by $\mu_c\simeq1.888$\footnote{When $\zeta$ is beyond
    a certain value, the superconducting phase transition will become a first order one. In that case, the critical
    chemical will depend on $\zeta$. }.
\begin{figure}[h]
\centering
\includegraphics[scale=0.9]{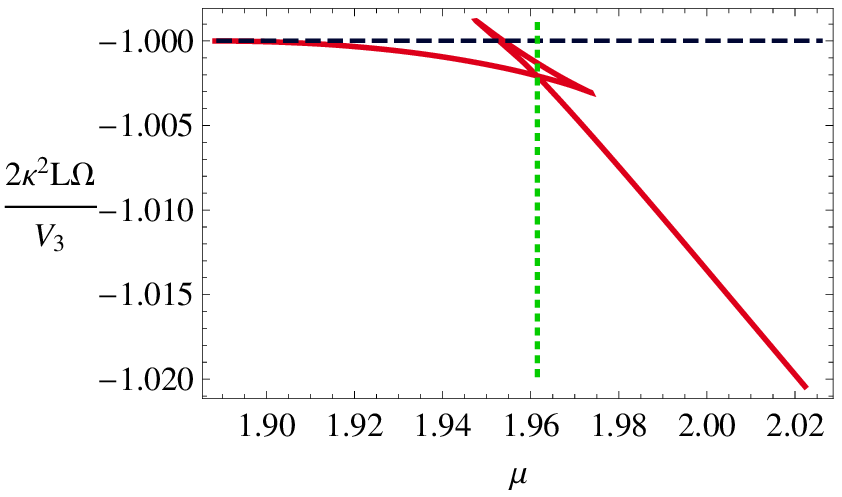}\ \ \ \
\includegraphics[scale=0.9]{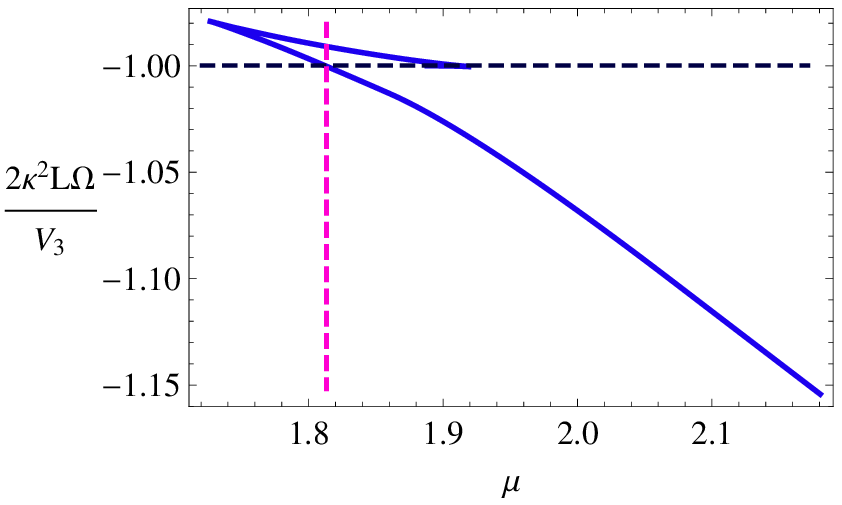} \caption{\label{smfree} The grand
potentials
of the soliton with scalar hair (solid) and the soliton without scalar hair (dashed black)
with respect to chemical potential for $\zeta=0.65$ (left plot) and $\zeta=1.6$ (right plot),
respectively. $\Gamma$ is scaled to be $\pi L$ and $\beta=0.25$. The left plot has a second order
phase transition at $\mu\simeq1.888$, but a discontinuity at $\mu_0\simeq1.962$ (denoted by dotted
 vertical green line) in superconducting phase. The right plot shows the typical first order
 phase transition and the critical chemical potential denoted by vertical dashed pink
 line is $\mu_c\simeq1.813$.}
\end{figure}

As we increase $\zeta$ to $\zeta\simeq0.5$, an additional complication appears. The condensate versus
chemical potential does not have a monotonic behavior (see the red curve in Figure~\eqref{smcondensate}).
As we can see clearly in the left plot of Figure~\eqref{smfree}, the grand potential
with respect to chemical potential develops a ``swallow tail", which is a typical signal
for a first order phase transition. It implies that there is a new phase transition within
the superconducting phase. If we continue to increase $\zeta$, the insulator/superconductor
phase transition will become first order when $\zeta>1$, which can be seen in
the right plot of Figure~\eqref{smfree}.\footnote{It is worth pointing out here that if the back
reaction parameter $\beta$ is larger enough, the transition will always be first order,
no matter how small the value of $\zeta$ is chosen. However, we do not consider the
case in this paper.} This is qualitatively similar to the
case in Ref.~\cite{Horowitz:2010jq} where the effect of back reaction on the insulator/superconductor was studied
in the case with vanishing $\zeta$.
The condensate becomes larger with the increase of $\zeta$, which means that the scalar hair will
be more difficult to be formed in the AdS soliton background with larger $\zeta$.

\section{Entanglement Entropy}
\label{sect:smhee}

Now let us begin to study the behavior of entanglement entropy in
this holographic superconductor model. The holographic method to
calculate entanglement entropy is as follows. Consider a strongly
coupled field theory with gravity dual,  the entanglement entropy of
subsystem $\mathcal{A}$ with its complement is given by searching
for the minimal area surface $\gamma_\mathcal{A}$ extended into the
bulk with the same boundary $\partial\mathcal{A}$ of $\mathcal{A}$.
Then the entanglement entropy of $\mathcal{A}$ with its complement
is given by the ``area law"~\cite{Ryu:2006bv}
\begin{equation}\label{law}
S_\mathcal{A}=\frac{2\pi}{\kappa^2} Area(\gamma_\mathcal{A}),
\end{equation}
where $\kappa$ is the gravitational constant.

We consider the subsystem ${\cal A}$ with a straight strip geometry
with a finite width $\ell$ along the $x$ direction,  along the
$\eta$ direction with a period $\Gamma$, but infinitely extending
along the $y$ direction.  The subsystem $\mathcal{A}$ sites on the
slice $r=\frac{1}{\epsilon}$, where $\epsilon\rightarrow0$ is the UV
cutoff.  The holographic dual surface $\gamma_A$ is defined as a
three-dimensional surface
\begin{equation}\label{embed}
t=0,\ \ r=r(x),\ \ -\frac{R}{2}<y<\frac{R}{2}\ (R\rightarrow\infty),\ \ 0\leq\eta\leq\Gamma.
\end{equation}
We are first interested in the case that the surface is smooth. The
holographic surface $\gamma_A$ starts from $x=\frac{\ell}{2}$ at
$r=\frac{1}{\epsilon}$, extends into the bulk until it reaches
$r=r_*$, then returns back to the AdS boundary
$r=\frac{1}{\epsilon}$ at $x=-\frac{\ell}{2}$.
\begin{figure}[h]
\centering
\includegraphics[scale=0.8]{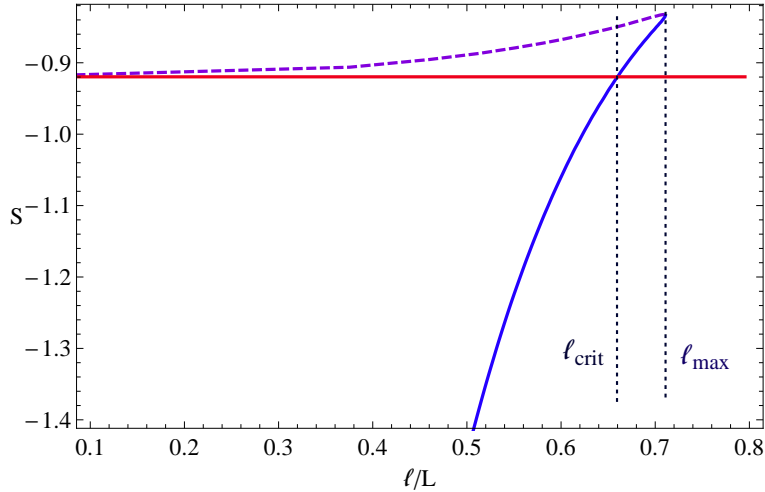}
\caption{\label{entropywidth} The entanglement entropy as a function of
strip width $\ell$ for $\beta=0.25$, $\zeta=0.65$ and $\mu\simeq2.276$. The dashed purple and solid blue
curves come from the connected configuration, while the solid red one
comes from the disconnected configuration. The lowest curve is physically
favored compared with others. In this figure $\ell_{crit}\simeq 0.66 L$ and $\ell_{max}\simeq 0.71 L$.}
\end{figure}
In this case,  the entanglement entropy is given by
\begin{equation}\label{entropy3}
S_\mathcal{A}^{connect}=\frac{4\pi L}{\kappa^2}R\Gamma \int_{r_*}^{\frac{1}{\epsilon}}
\frac{r^4\sqrt{g(r)}e^{-\chi(r)}}{\sqrt{r^6 g(r)e^{-\chi(r)}-r_*^6 g(r_*)e^{-\chi(r_*)}}}dr=\frac{2\pi L}
{\kappa^2}R\Gamma(\frac{1}{\epsilon^2}+S^{con}),
\end{equation}
where the UV cutoff $1/\epsilon$ has been taken into consideration.
The width $\ell$ of the subsystem $\mathcal{A}$ and $r_*$ are
related  by
\begin{equation}\label{width2}
\frac{\ell}{2}=\int_{r_*}^{\frac{1}{\epsilon}}
\frac{L}{r^2\sqrt{g(r)(\frac{r^6 g(r)e^{-\chi(r)}}{r_*^6 g(r_*)e^{-\chi(r_*)}}-1)}}dr.
\end{equation}
In fact, there are two solutions for the connected configuration.
See Figure~\eqref{entropywidth}. In addition, there is also a
disconnected configuration describing two separated surfaces located
at $x=\pm\frac{\ell}{2}$ respectively and extending to the bulk and
reaching at the tip of the soliton geometry. The entropy for this
disconnected geometry is independent of $\ell$, and given by
\begin{equation}\label{entropy2}
S_\mathcal{A}^{disconnect}=\frac{4\pi L}{\kappa^2}R\Gamma \int_{r_0}^{\frac{1}{\epsilon}}
 r e^{-\frac{\chi(r)}{2}}dr=\frac{2\pi L}{\kappa^2}R\Gamma(\frac{1}{\epsilon^2}+S^{discon}).
\end{equation}

We find that the entanglement entropy with respect to strip width $\ell$ behaves
quite similar for different choice of parameters, i.e., $\beta$, $\zeta$ and $\mu$.
 We draw Figure~\eqref{entropywidth} with $\beta=0.25$, $\zeta=0.65$ and $\mu\simeq2.276$ as a concrete example.
 The connected configuration does not exist when $\ell>\ell_{max}$. The physical solution comes from the trivial
 disconnected geometry and the entropy is independent of strip width. On the contrary, there are three different
 branches when $\ell<\ell_{max}$, i.e., the upper branch (dashed purple), the lower branch (solid blue) and the
 middle branch (solid red). The first two curves correspond to the connected surface, while the third curve comes
 from the disconnected one. We trace over the physical entropy by always choosing the lowest branch. As we can see
 from Figure~\eqref{entropywidth}, the lower branch (solid blue) is finally favored as we decrease $\ell$. Hence,
  there is a critical value $\ell_{crit}$ below which the lower branch is physically favored. Thus, as we
   change $\ell$, a phase transition occurs at $\ell_{crit}$, which is just the so
   called ``confinement/deconfinement" phase transition~\cite{con1,con2,Myers:2012ed}. To
    be precise, for $\ell<\ell_{crit}$, the entanglement entropy comes from the connected surface
    and exhibits non-trivial dependence on $\ell$, which describes a ``deconfiement" phase.
    For $\ell>\ell_{crit}$, the entropy is dominated by the disconnected configuration and is
    $\ell$ independent, which indicates a ``confinement" phase.

\begin{figure}[h]
\centering
\includegraphics[scale=0.9]{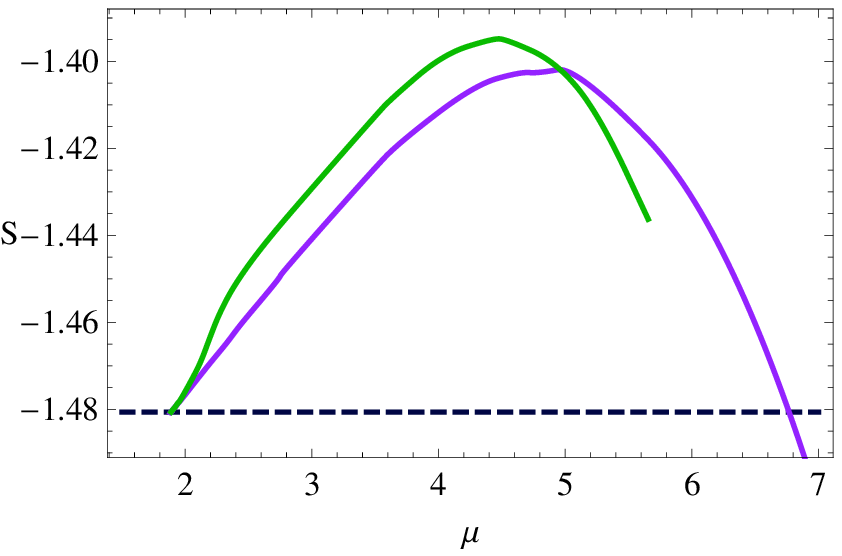}\ \ \ \
\includegraphics[scale=0.9]{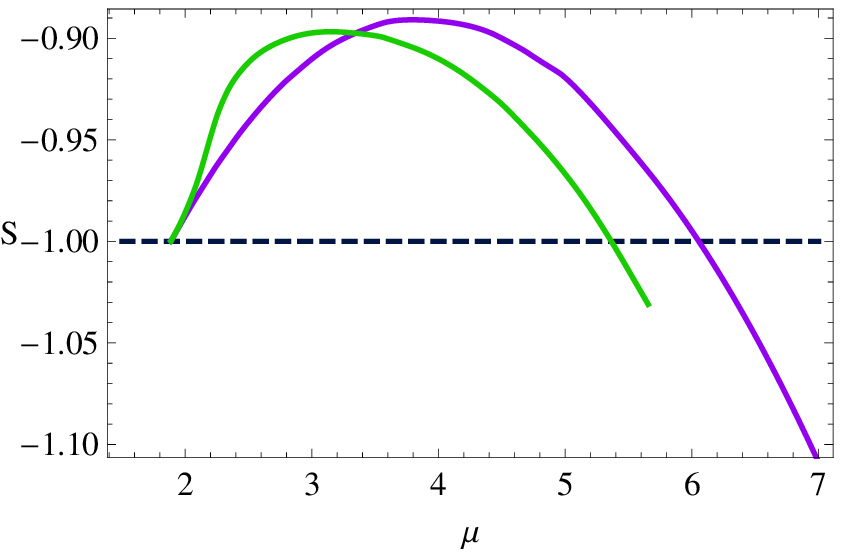} \caption{\label{smhee1} The entanglement entropy
 as a function of chemical potential for $\zeta=0$ (purple curves) and $\zeta=0.2$
 (green curves) at fixed $\ell/L=0.5$ (within ``deconfining phase", left plot) and
  $\ell/L\rightarrow\infty$ (within ``confining phase", right plot), respectively.
   The transition from insulator to superconductor is second order here. In both two plots,
   the solid curves come from superconducting phase, while the dashed black lines come from insulator phase.}
\end{figure}

\begin{figure}[h]
\centering
\includegraphics[scale=0.9]{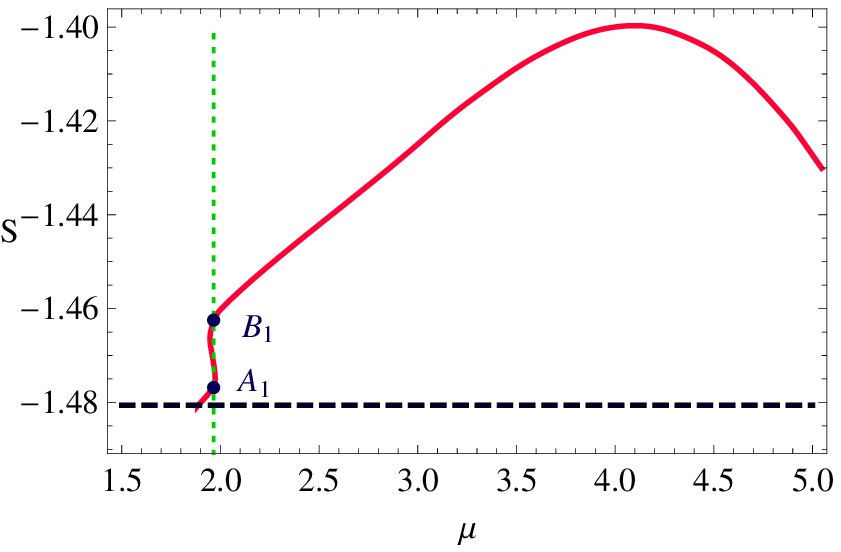}\ \ \ \
\includegraphics[scale=0.9]{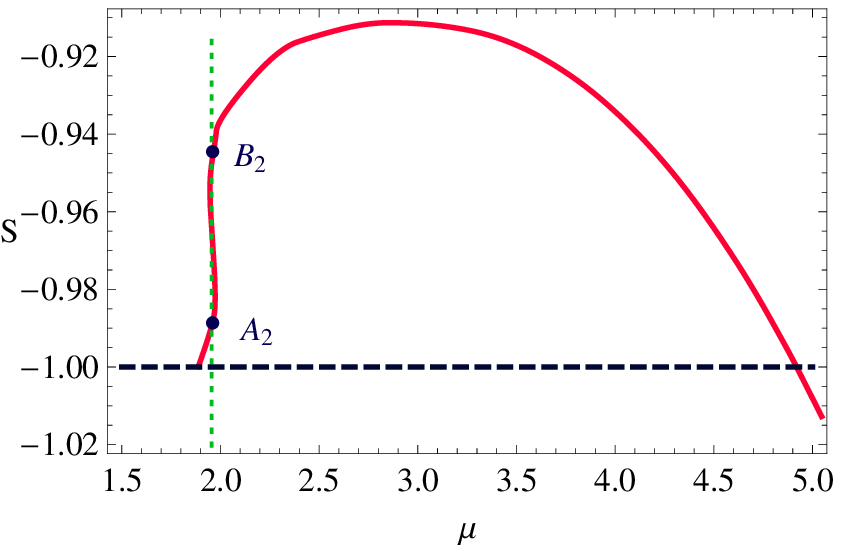} \caption{\label{smhee2} The entanglement entropy as a function of
 chemical potential for $\zeta=0.65$ at fixed $\ell/L=0.5$ (within ``deconfinement phase", left plot)
  and $\ell/L\rightarrow\infty$ (within ``confinement phase", right plot), respectively. The transition
   from insulator to superconductor is second order, but a discontinuity at
    $\mu_0\simeq1.962$ (denoted by dotted vertical line) within the superconducting phase. In both
    two plots, the solid red curves are from superconducting phase, while the dashed black lines are
    from insulator phase. Trace the physical curve by choosing the dashed black line for $\mu<\mu_c$,
    then choosing the red curve which has a jump from point A to point B at $\mu_0\simeq1.962$.}
\end{figure}

To summarize, there are totally four ``phases" in the dual boundary
field theory, i.e., the insulator phase, superconductor phase, and
their corresponding confinement/deconfinement phases. These phases
are characterized by the chemical potential $\mu$ and strip width
$\ell$. In particular, the strip width controls the
``confinement/deconfinement" phase transition\footnote{Strictly
speaking, the term phase transition here is inappropriate since the
system itself, i.e., the state of the boundary field theory, does
not change at all as one changes $\ell$. However, this observed
behavior here is quite similar to the one in the thermodynamic
confinement/deconfinement phase transitions and therefore  we follow
Refs.~\cite{con1,con2,Myers:2012ed} and adopt the terminology
``phase transition" to describe this behavior.}.

It is instructive to study how the entanglement entropy changes with
chemical potential
 by fixing strip width $\ell$.
 We first focus on the second order phase transition case, which is presented in
  Figure~\eqref{smhee1}. We can see that, at the beginning of the phase transition,
  the entanglement entropy  increases continuously with chemical
  potential both in ``deconfinement phase" and ``confinement phase" and reaches its
   maximum at a certain chemical potential $\mu_{max}$, then it decrease monotonically.
   Furthermore, the entanglement entropy is continuous at critical chemical potential $\mu_c$,
   but its slop has a discontinuous change at $\mu_c$. The behavior
   of the entanglement entropy across the phase transition point
   indicates that the transition is a second order one.
\begin{figure}[h]
\centering
\includegraphics[scale=0.9]{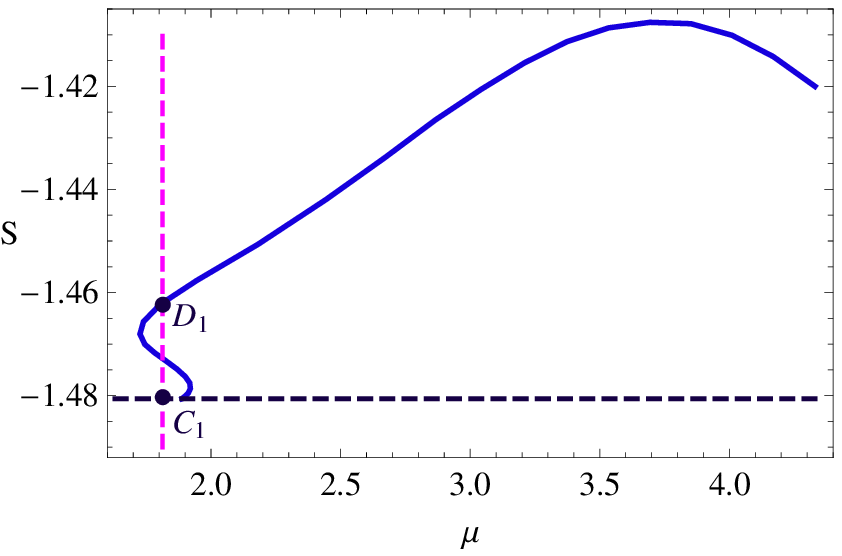}\ \ \ \
\includegraphics[scale=0.9]{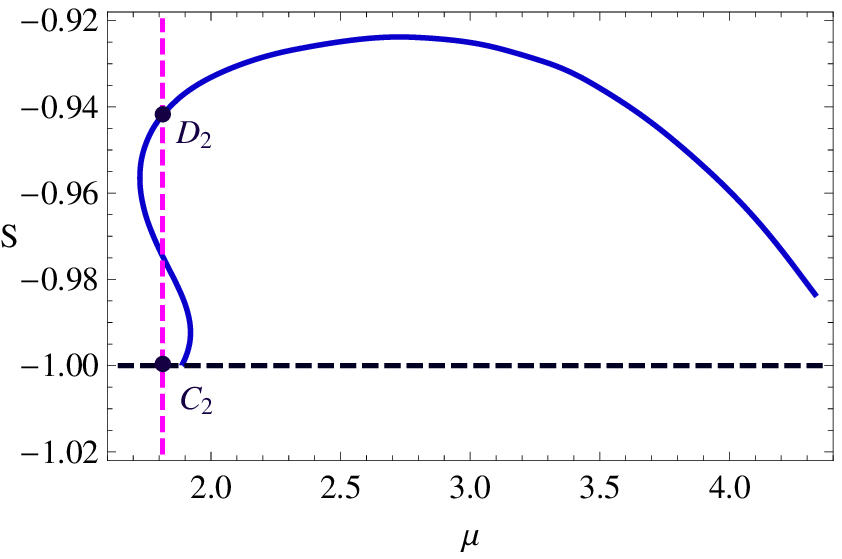} \caption{\label{smhee3} The entanglement entropy as a function
of chemical potential for $\zeta=1.6$ at fixed $\ell/L=0.5$ (within
``deconfinement phase", left plot) and $\ell/L\rightarrow\infty$
(within ``confinement phase", right plot), respectively. The
transition is first order and the critical chemical potential
denoted by vertical dashed pink line
 is $\mu_c\simeq1.813$. In both two plots, the solid blue curves are from superconducting phase,
 while the dashed black lines are from insulator phase. Trace the physical curve by choosing
 the dashed black line for $\mu<\mu_c$, and for $\mu>\mu_c$, choosing the blue curve starting from point D.}
\end{figure}

When $0.5<\zeta<1$, a jump in the condensate (the red curve in
Figure~\eqref{smcondensate}) and a ``swallow tail" in the grand
potential (see the left plot in Figure~\eqref{smfree}) appear,
although the phase transition is still a second order one at
$\mu_c$. Interestingly, we can also see a jump in
Figure~\eqref{smhee2} which shows the entanglement entropy as a
function of chemical potential at fixed strip width. Tracing the
physical curve, we find that the entropy increases at the beginning
of the transition, then there is a sudden jump at the chemical
potential $\mu_0\simeq1.962$. The sudden jump of entropy indicates a
first order phase transition there.

For sufficiently large $\zeta$, the insulator/superconductor transition becomes first order.
The entanglement entropy with respect to chemical potential is presented in Figure~\eqref{smhee3}.
Comparing with Figure~\eqref{smhee1} and~\eqref{smhee2}, we can see a dramatic change in the first
 order case. Although the entropy in superconducting phase also behaves non-monotonically with respect to $\mu$,
  the entropy as well as its slop at the critical point have a discontinuous
  jump. Once again, it is a signal of a first order transition.

\begin{figure}[h]
\centering
\includegraphics[scale=0.9]{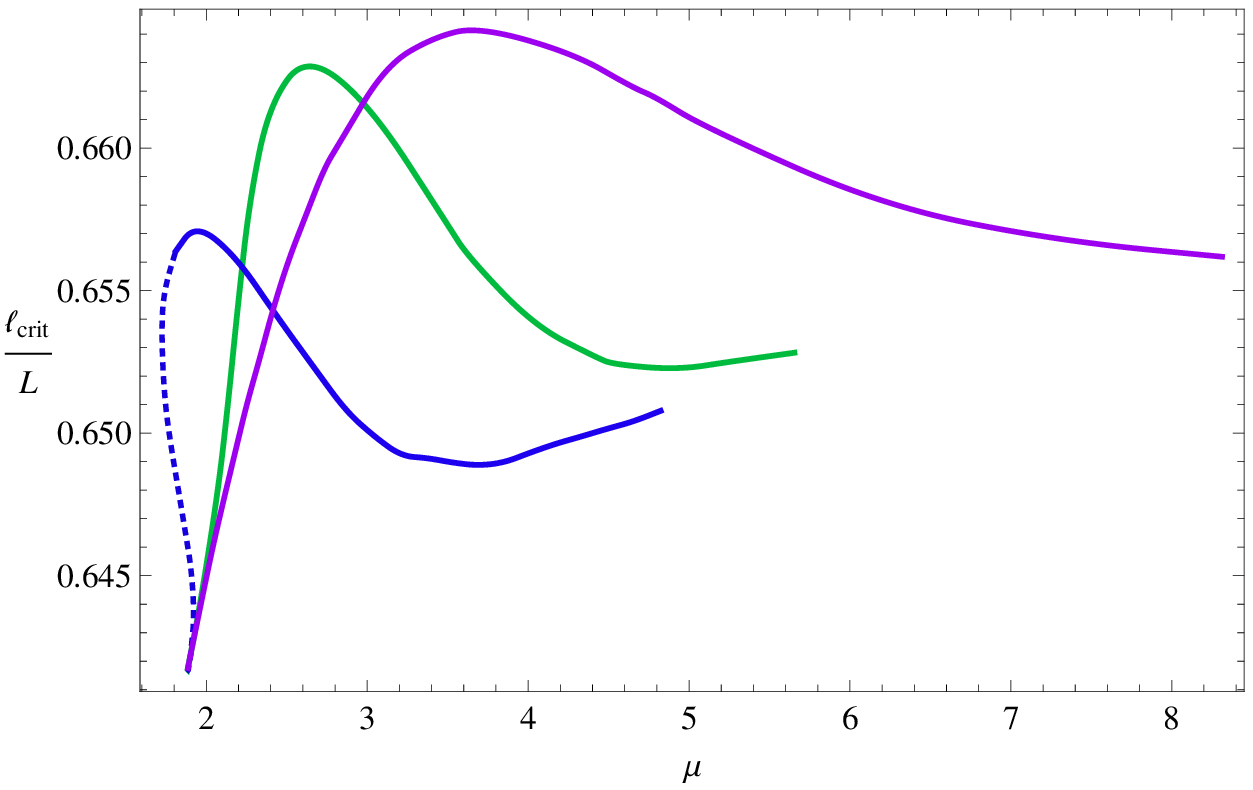}
\caption{\label{heelc} The critical length as a function of chemical
potential in superconducting phase, where the purple curve (up) is
for $\zeta=0$, the green curve (middle) is for $\zeta=0.2$ and the
blue curve (down) is for $\zeta=1.6$. The dotted part of the line
for $\zeta=1.6$ (blue) is not thermodynamically favored, trace the physical curve by always choosing the
solid line. Note that the curve for $\zeta=0$ (purple) will continue to increases
large chemical potential as other curves.}
\end{figure}

In Figure~\eqref{heelc}, we plot the critical length $\ell_{crit}$
of the ``confinement/deconfinement" phase transition with respect to
the chemical potential in the superconducting phase. For small
$\zeta$, $\ell_{crit}$ first increases and forms a peak, then it will
increase continuously for large chemical potential. As we increase
$\zeta$, the position of the peak moves toward to the region with
small chemical potential and will be finally cut off since the
transition is first order for large $\zeta$, as we can see that the
dotted part of the blue curve in Figure~\eqref{heelc} is not
thermodynamically favored. Here an remarkable point is that the
critical length as a function of the chemical potential is not
monotonic. This is of course related to the fact that the
entanglement entropy is not a monotonic function of the chemical
potential.

We can see from Figures~\eqref{smhee1},~\eqref{smhee2} and
\eqref{smhee3} that although the phase structures are different in
three cases, the entanglement entropy with respect to chemical
potential are always non-monotonic in superconducting phase.  The
discontinuity or jump at critical point indicates some kind of
significant reorganization of the degrees of freedom of the system,
since
 new degrees of freedom are expected to emerge in new phase. The
 non-monotonic behavior of the entanglement entropy is interesting.
 To further study its implication, in the next section, we will
 discuss the behavior of another non-local quantity, Wilson loop, in
 the superconducting phase.

\section{Wilson Loop}
\label{sect:wilson}

 From the VEV of Wilson loop one can extract the
interaction potential between a quark-antiquark pair in gauge field
theory. Therefore the VEV of Wilson loop is
   thought of as a proper quantity to describe the confinement/deconfinement
   phase transition of gauge field theory. In the AdS/CFT
   correspondence, a proposal for the VEV of Wilson loop is
   presented in
   Refs.~\cite{Maldacena:1998im,Rey:1998bq,Polyakov:1997tj}. In this section, following
   the proposal we will
   discuss the behavior of the VEV of Wilson
loop  in the holographic insulator/superconductor phase transition.
We will consider two kinds of Wilson loop: One is a temporal Wilson
loop; the other is a spatial Wilson loop. In both cases, we will
show the behavior of the interaction potential with respect to the
distance $\ell$ between quark and antiquark and to the background
chemical potential $\mu$.

\subsection{Temporal Wilson Loop}
\label{sect:wilsonxt}

 In $SU(N)$ gauge theory, the interaction potential
 for heavy quark-antiquark ($Q{\bar Q}$) pair can be calculated from the Wilson loop
\begin{equation}
W[C]=\frac{1}{N} Tr P \exp[i \oint_{C} A_\mu dx^\mu],
\label{Wilson-loop-formula}
\end{equation}
where $A_{\mu}$ is the gauge field, the trace is over the fundamental representation, $P$ stands
for path ordering. $C$ denotes a closed loop in space-time, which is a rectangle with one direction
along the temporal direction of length $T$ and the other spatial direction of length $\ell$. The Wilson
loop describes the creation of a $Q{\bar Q}$ pair with distance $\ell$ at some time $t_0=0$ and the annihilation
of this pair at time $t=T$. For $T\to\infty$, the VEV of the Wilson loop goes
 as $\langle W(C)\rangle\propto e^{-T V_{Q\bar Q}}$. In terms of the
 AdS/CFT dictionary, the VEV of the Wilson loop in four dimensions should be equal to the string partition
  function on the curved space, with the string world sheet ending on the contour $C$ at the boundary
  of the curved space~\cite{Maldacena:1998im}
\begin{equation}\label{dictionary}
\langle W^{4d}[C]\rangle=Z_{string}^{5d}[C]\simeq e^{-S_{x-t}[C]}
\,\ ,
\end{equation}
where $S_{x-t}$ is the classical world sheet  action of the
Nambu-Goto form
\begin{equation} \label{S-NG-HQ}
S_{x-t}=\frac{1}{2\pi\alpha'}\int d \tau d\sigma\sqrt{{\rm Det}
\chi_{a b}},
\end{equation}
and $\alpha'$ is the string tension  with dimension
${[energy]}^{-2}$, and $\chi_{a b}$ is the induced world sheet
metric with $a,b$ the indices in the ($\tau=t,\sigma=x)$ coordinates
on the world sheet. In this subsection, we follow the standard
procedure \cite{Maldacena:1998im,Rey:1998bq,Polyakov:1997tj} to
extract the static heavy quark potential $V_{Q{\bar Q}}$ in the
general metric background (\ref{metric2}). To do this, as in
Ref.~\cite{Cai:2012xh}, we should consider two kinds of string
configurations: one is a connected configuration and the other is a
disconnected one.

For the connected configuration, the string starts at the AdS
boundary $r=1/\epsilon$ and $x=\ell/2$, goes into the bulk, turns
abound at $r=r_*$ and $x=0$, and finally reaches at $r=1/\epsilon$
and $x=-\ell/2$. For the disconnected configuration, two straight
strings start at $r=1/\epsilon$ with $x= \pm \ell/2$, respectively
and end at the tip of the background soliton, $r=r_0$. Here
$1/\epsilon$ is the UV cutoff and $\ell$ is the distance between two
quarks.

We now take the gauge $t =\tau$ and $x(r)=\sigma$.
 Then the induced metric $\chi_{a b}$ can
be expressed as
\begin{equation}\label{inducetx}
ds^2=\chi_{a b}dx^a
dx^b=g_{\mu\nu}\frac{dx^\mu}{dx^a}\frac{dx^\nu}{dx^b} dx^a dx^b
=[\frac{L^2}{r^2
g(r)}(\frac{dr}{d\sigma})^2+r^2]d\sigma^2-r^2f(r)d\tau^2.
\end{equation}
Thus, we need to minimize the following Nambu-Goto action functional
\begin{equation}\label{nambutx}
S_{x-t}^{con}[r]=\frac{1}{2\pi\alpha'}\int_0^T\int_{-\frac{\ell}{2}}^{\frac{\ell}{2}}
d\tau d\sigma \sqrt{(\frac{L^2}{r^2 g(r)}(\frac{dr}{d\sigma})^2+r^2)r^2 f(r)}.
\end{equation}
\begin{figure}[h]
\centering
\includegraphics[scale=0.9]{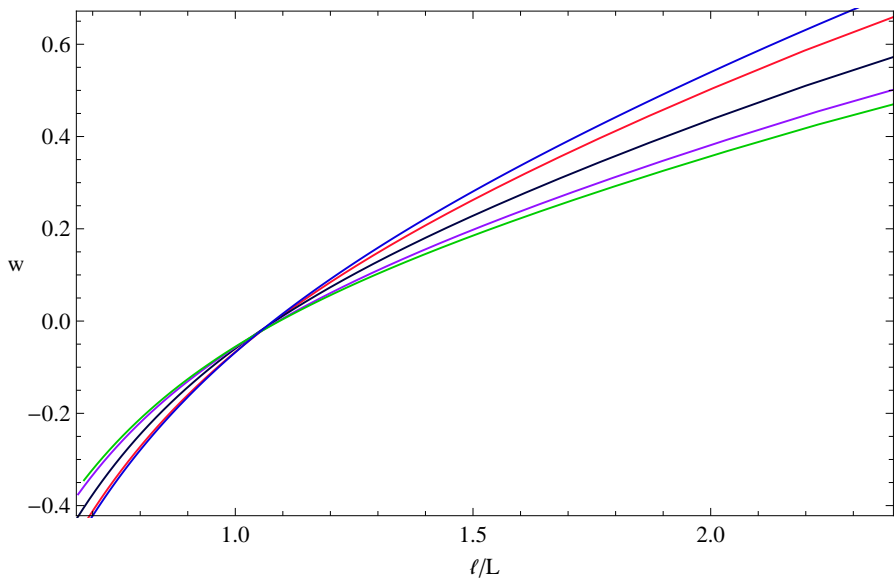}
\caption{\label{xtlength} The heavy quark potential as a function of
distance $\ell$ in the case $\zeta=0.2$. In the right part of the
plot, lines from the top to down are for $\mu\leq\mu_c$ (blue),
$\mu\simeq2.134$ (red), $\mu\simeq3.025$ (black), $\mu\simeq4.920$ (purple) and
$\mu\simeq5.655$ (green), respectively.}
\end{figure}

The integrand can be regarded as a Lagrangian and $\sigma$ as time.
Taking advantage of the fact that the Hamiltonian is conserved, we
can deduce the equation of motion
\begin{equation}\label{eomtx}
\frac{dr}{d\sigma}=\pm\frac{r^2\sqrt{g(r)}}{L}\sqrt{\frac{r^4f(r)}{r_*^4f(r_*)}-1},
\end{equation}
where we have used the condition that the geometry is smooth at
$r(x=0)=r_*$, i.e., $dr/d\sigma|_{r=r_*}=0$.
Substituting~\eqref{eomtx} to~\eqref{nambutx}, we obtain the
on-shell world sheet action in the connected case
\begin{equation}\label{wilsontx}
S_{x-t}^{con}=\frac{2T}{2\pi\alpha'} \int_{r_*}^{\frac{1}{\epsilon}} dr \sqrt{\frac{f(r)}{g(r)}}
\frac{L}{1-\sqrt{\frac{r^4f(r)}{r_*^4f(r_*)}}}=\frac{2TL}{2\pi\alpha'}(\frac{1}{\epsilon}+w^{con}),
\end{equation}
where the UV cutoff $\epsilon$ has been taken into consideration and
$w^{con}$ is a finite
 part. The distance $\ell$ is
related to $r_*$ by
\begin{equation}\label{widthtx}
\frac{\ell}{2}=\int_{r_*}^{\frac{1}{\epsilon}} dr \frac{1}{r^2\sqrt{g(r)}}\frac{L}
{\sqrt{\frac{r^4f(r)}{r_*^4f(r_*)}-1}}.
\end{equation}
 For the disconnected configuration, with using
the gauge $\tau=t$ and $\sigma=r$ and $x'(r)=0$, we obtain the regularized on-shell world
 sheet action as follows.
\begin{equation}\label{wilsontxdis}
S_{x-t}^{discon}=\frac{2T}{2\pi\alpha'}
\int_{r_0}^{\frac{1}{\epsilon}} dr
L\sqrt{\frac{f(r)}{g(r)}}=\frac{2TL}{2\pi\alpha'}(\frac{1}{\epsilon}+w^{discon}).
\end{equation}
One can see that $S_{x-t}^{con}$ and $S_{x-t}^{discon}$  have the
same UV behavior. The divergence in (\ref{wilsontxdis}) just
manifests the infinite mass of heavy quarks.  The divergence in
(\ref{wilsontx}) can be subtracted by considering the infinite mass
of quarks in (\ref{wilsontxdis}). And thus we can obtain the heavy
quark potential $w\equiv w^{con}-w^{discon}$ as a function of the
distance $\ell$. We find that the behavior of the heavy quark
potential is quite similar for different parameters, i.e., $\beta$,
$\zeta$ and $\mu$. We present Figure~\eqref{xtlength} with
$\zeta=0.2$ as a concrete example.  One can see from the figure that
in all cases, the heavy quark potential goes roughly linear with the
distance in large $\ell$ region, which implies the dual field theory
is a confinement phase, while in the small $\ell$ region, the
potential shows a Coulomb potential behavior, which means that the
dual field theory is in a deconfinement phase in the small $\ell$
region. There is an associated confinement/deconfinement phase
transition when one changes the distance $\ell$. This feature is
consistent with the one shown in Figure~\eqref{entropywidth} through
the entanglement entropy calculation.  Here the
confinement/deconfinement transition point can be identified with
the point where the potential vanishes.
\begin{figure}[h]
\centering
\includegraphics[scale=0.9]{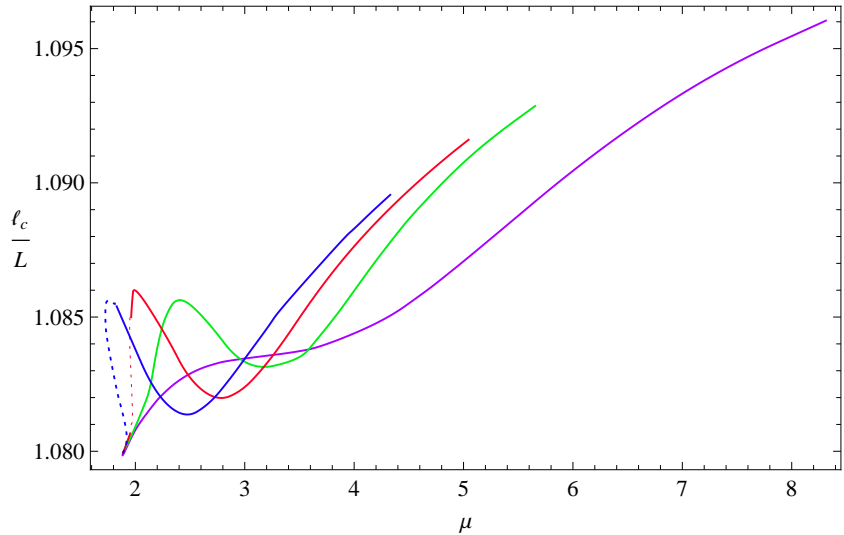}
\caption{\label{xtlc} The critical distance $\ell_c$ as a function
of chemical potential. Lines from right to left are for $\zeta=0$
(purple), $\zeta=0.2$ (green), $\zeta=0.65$ (red)
 and $\zeta=1.6$ (blue), respectively. The dotted part of the lines for $\zeta=0.65$ (red)
 and $\zeta=1.6$ (blue) are not thermodynamically favored, trace the physical curves by always choosing the solid lines.}
\end{figure}

\begin{figure}[h]
\centering
\includegraphics[scale=0.6]{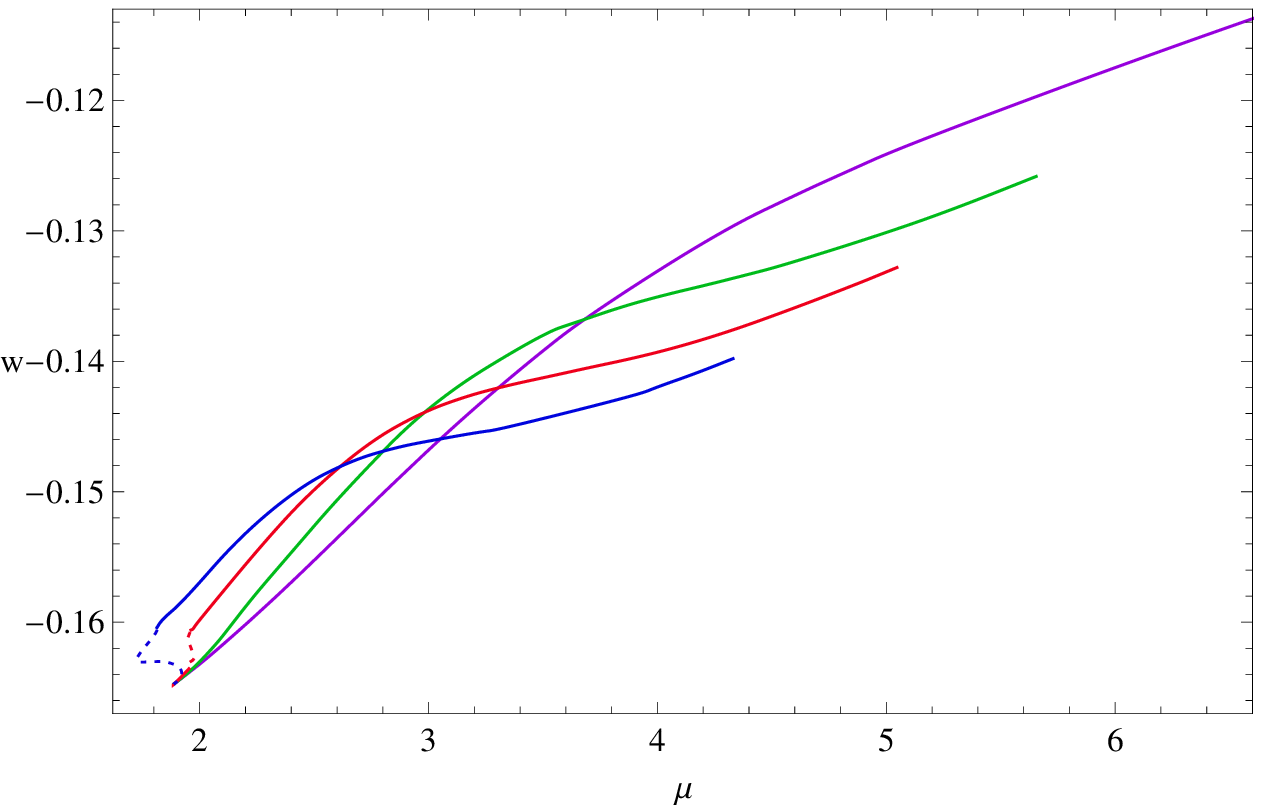} \ \ \ \
\includegraphics[scale=0.6]{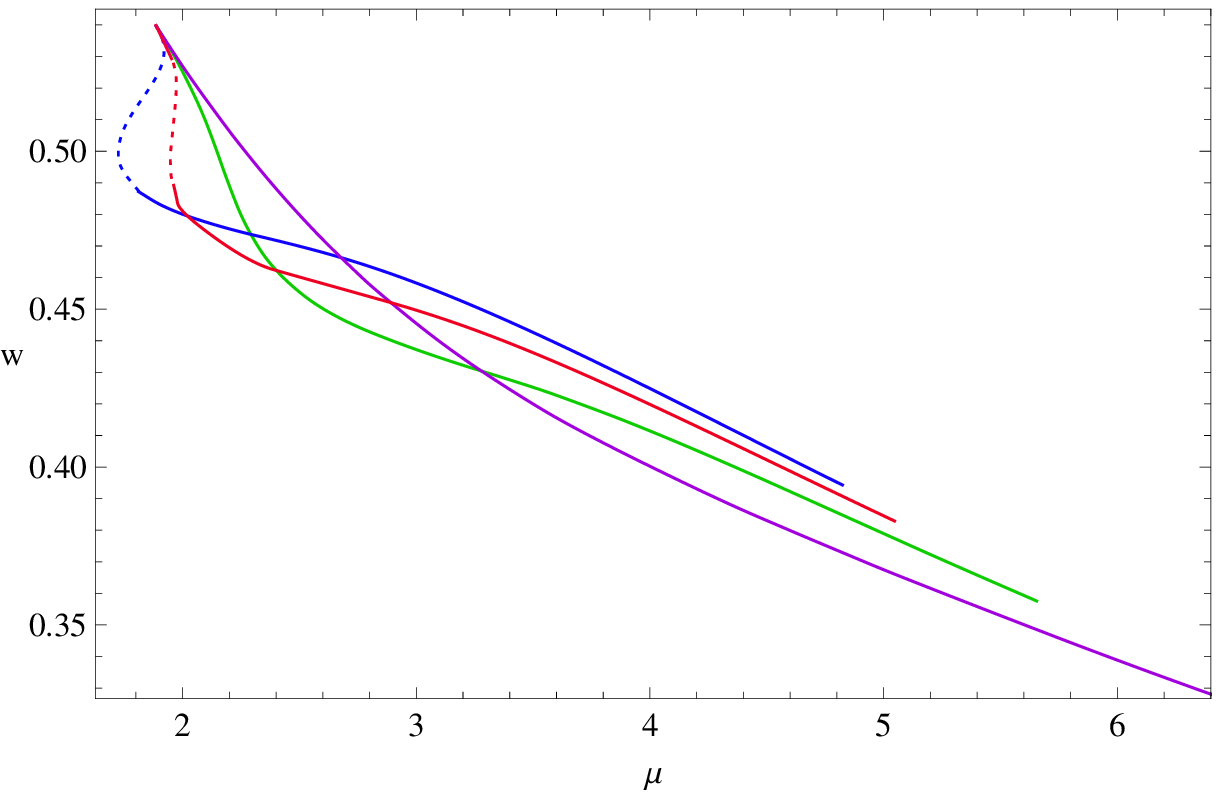}
\caption{\label{xtlittle} The  heavy quark potential as a function
of chemical potential at fixed $\ell/L=0.9$ (left plot) and
$\ell/L=2$ (right plot). Lines are for
 $\zeta=0$ (purple), $\zeta=0.2$ (green), $\zeta=0.65$ (red) and $\zeta=1.6$ (blue),
 respectively. The dotted part of the
lines for $\zeta=0.65$ (red) and $\zeta=1.6$ (blue) are not
thermodynamically favored,
 trace the physical curves by always choosing the solid lines.}
\end{figure}

In Figure~\eqref{xtlc} we plot the relation of the critical distance
$\ell_c$ with respect to chemical potential. One can see that
$\ell_c$ increases monotonically when $\zeta$ is very small, then a
peak emerges and
 the position of the peak moves to the left as we increase $\zeta$, finally, the peak is cut off for very
 large $\zeta$ because there is a jump at $\mu_c$ for the first order phase
 transition. One interesting point is that the critical distance is
 not always a monotonic function of chemical potential. We think that this arises due to the different
  behavior of the heavy quark
 potential in the small and large $\ell$ regions.
In Figure~\eqref{xtlittle}, we plot the heavy quark potentials
versus chemical potential in the deconfinement phase and confinement
phase, respectively. We can see clearly that the behavior of the
heavy quark potential is quite different in two cases.

\subsection{Spatial Wilson Loop}
\label{sect:wilsonxy}

\begin{figure}[h]
\centering
\includegraphics[scale=0.9]{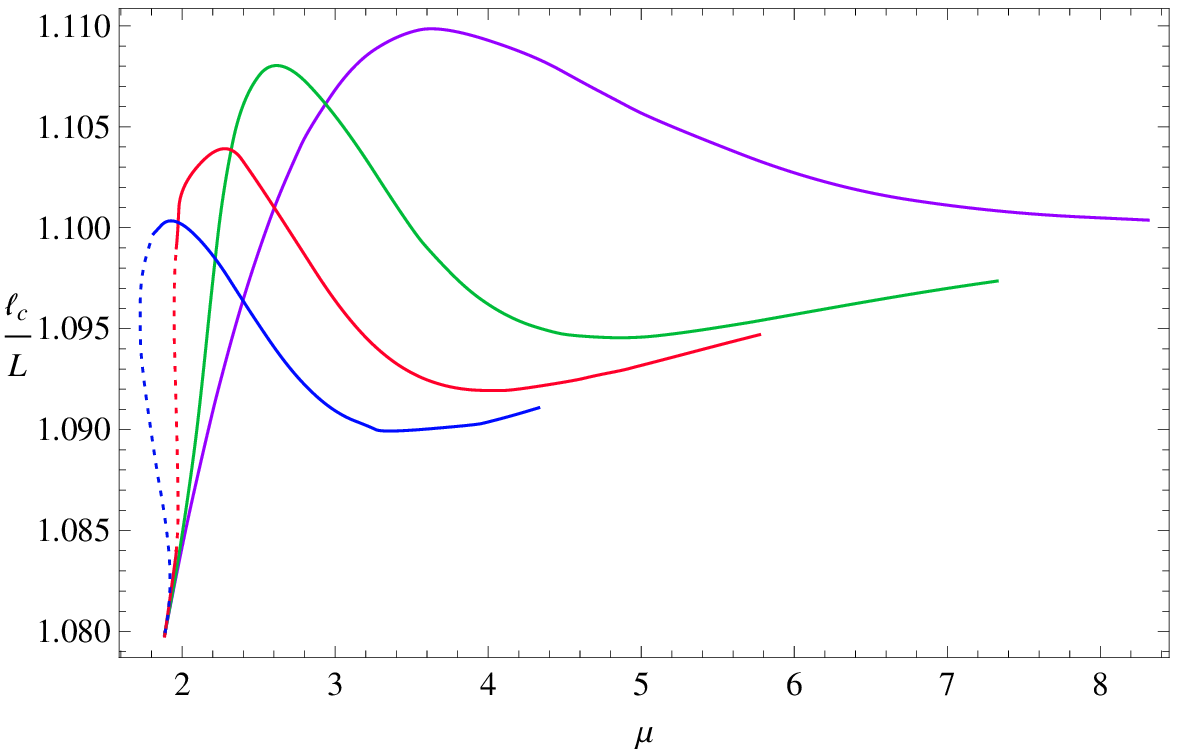}
\caption{\label{xylc} The critical distance $\ell_c$ as a function
of chemical potential in the case of spatial Wilson loop. Lines from
right to left are for $\zeta=0$ (purple), $\zeta=0.2$ (green),
$\zeta=0.65$ (red) and $\zeta=1.6$ (blue), respectively. The dotted
part of the lines for $\zeta=0.65$ (red) and $\zeta=1.6$ (blue) are
not thermodynamically favored, trace the physical curves by always
choosing the solid lines. Note that the curve for $\zeta=0$ (purple)
will continue to increases for sufficiently large chemical potential
as other curves.}
\end{figure}

In terms of gauge/gravity duality, to calculate the VEV of a spatial
Wilson loop, we consider a rectangular loop $\mathcal{C}$ along two
spatial directions $(x, y)$~\cite{Andreev-T1}\cite{Li:2011hp}.
 The expectation value of the loop can be obtained
by the following AdS/CFT dictionary
\begin{equation}
\label{spatial-Wilson} V_s =\int D X e^{-S_{x-y}}.
\end{equation}
Here $X$ denotes the series of world sheet fields and $S_{x-y}$
stands for a world sheet action. Again, in the large N limit, the
saddle point approximation is valid, and we only need to take the
minimum value of the Euclidean action among the saddle points. Given
the background (\ref{metric2}), we can derive the expectation value
of the spatial Wilson loop by using the Nambu-Goto action for
$S_{x-y}$
\begin{eqnarray}
  S_{x-y} &=&  \frac{1}{2\pi
  \alpha'}\int d^2\eta \sqrt{{\rm Det}\chi_{ab}} \nonumber\\
  &=&{1\over 2\pi \alpha'} \int dx dy \sqrt{r^2(r^2+{L^2\over r^2}{r'(x)^2\over
  g(r)})},\label{worldsheetxy}
\end{eqnarray}
with $\alpha'$ the string tension and $\chi_{ab}$ the induced world sheet metric with $a,b$ the
 indices in the $(\eta^1=x,\eta^2=y)$ coordinates on the world sheet. We take one of the
 spatial direction $y$ goes to infinity. The quark and anti-quark are located at $x=\pm\frac{\ell}{2}$,
  respectively. As the case with the temporal Wilson loop, in this
  case, there are also two configurations. For the
 connected configuration, we can make use of $r'{\partial L\over r'}-L=-r_*^2$ and write down equation of motion for $r'(x)$
\begin{eqnarray}
{r^4\over \sqrt{r^2(r^2+ {L^2\over r^2}{r'(x)^2\over g(r)})}}=r_*^2.
\end{eqnarray}
Here we have used the following boundary condition $r(x=0)=r_*,
r'(x=0)=0$. We  put the solution of the equation of motion back to
the world sheet action (\ref{worldsheetxy}) and obtain
\begin{equation}\label{wilsonxy}
S_{x-y}^{con}=\frac{2Y}{2\pi\alpha'} \int_{r_*}^{\frac{1}{\epsilon}}dr
\frac{L}{\sqrt{g(r)}}\frac{r^2}{r^4-r_*^4}=\frac{2YL}{2\pi\alpha'}(\frac{1}{\epsilon}+w^{con}),
\end{equation}
where $Y$ denotes the separation of quark and antiquark in $y$
direction and the distance $\ell$ in $x$ direction is given by
\begin{equation}\label{widthxy}
\frac{\ell}{2}=\int_{r_*}^{\frac{1}{\epsilon}} dr \frac{L}{\sqrt{g(r)}}\frac{r_*^2}{r^2}\frac{1}{r^4-r_*^4}.
\end{equation}
Similar to the case for the temporal Wilson loop, we can obtain a
finite spatial heavy quark potential (pseudo potential) by
subtracting the mass of two heavy quarks. The latter is related to
the disconnected configuration.
 For the disconnected configuration, we  take the gauge $\eta^1=x$ and $\eta^2=y$ and
 $x'(r)=0$, and obtain the
 on-shell world sheet action
\begin{equation}\label{wilsonxydis}
S_{x-y}^{discon}=\frac{2Y}{2\pi\alpha'}
\int_{r_0}^{\frac{1}{\epsilon}} dr
\frac{L}{\sqrt{g(r)}}=\frac{2YL}{2\pi\alpha'}(\frac{1}{\epsilon}+w^{discon}).
\end{equation}
We find that the behavior of the pseudo potential $w\equiv
w^{con}-w^{discon}$ with respect to $\ell$ is very similar to the
one in the case of temporal Wilson loop. We therefore do not show
the behavior of the pseudo potential here. Instead, we plot in
Figure~\eqref{xylc} the critical distance $\ell_c$ with respect to
chemical potential. Here the critical distance $\ell_c$ is defined
by a vanishing pseudo potential as the case in the temporal Wilson
loop. We can see that $\ell_c$ is non-monotonic and a peak appears
even when $\zeta$ is vanishing, which is a little bit different from
the case in Figure~\eqref{xtlc}. For large chemical potential, the
critical distance increases monotonically. Comparing with
Figure~\eqref{heelc}, we find that the critical strip width
$\ell_{crit}$ from entanglement entropy and the critical distance
$\ell_c$ from the spatial Wilson loop are much qualitatively similar
with each other. This is expected since both quantities come from
the spatial sector of the background soliton solutions.

\begin{figure}[h]
\centering
\includegraphics[scale=0.6]{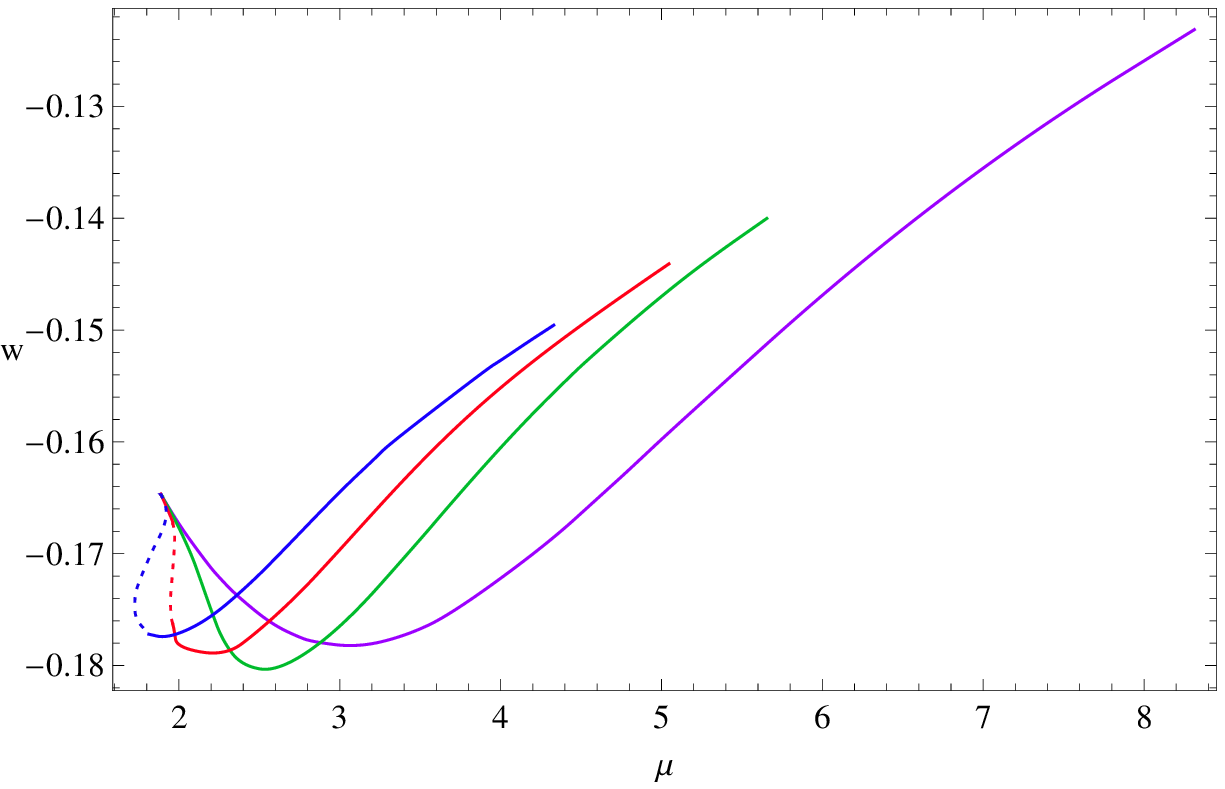}\ \ \
\includegraphics[scale=0.6]{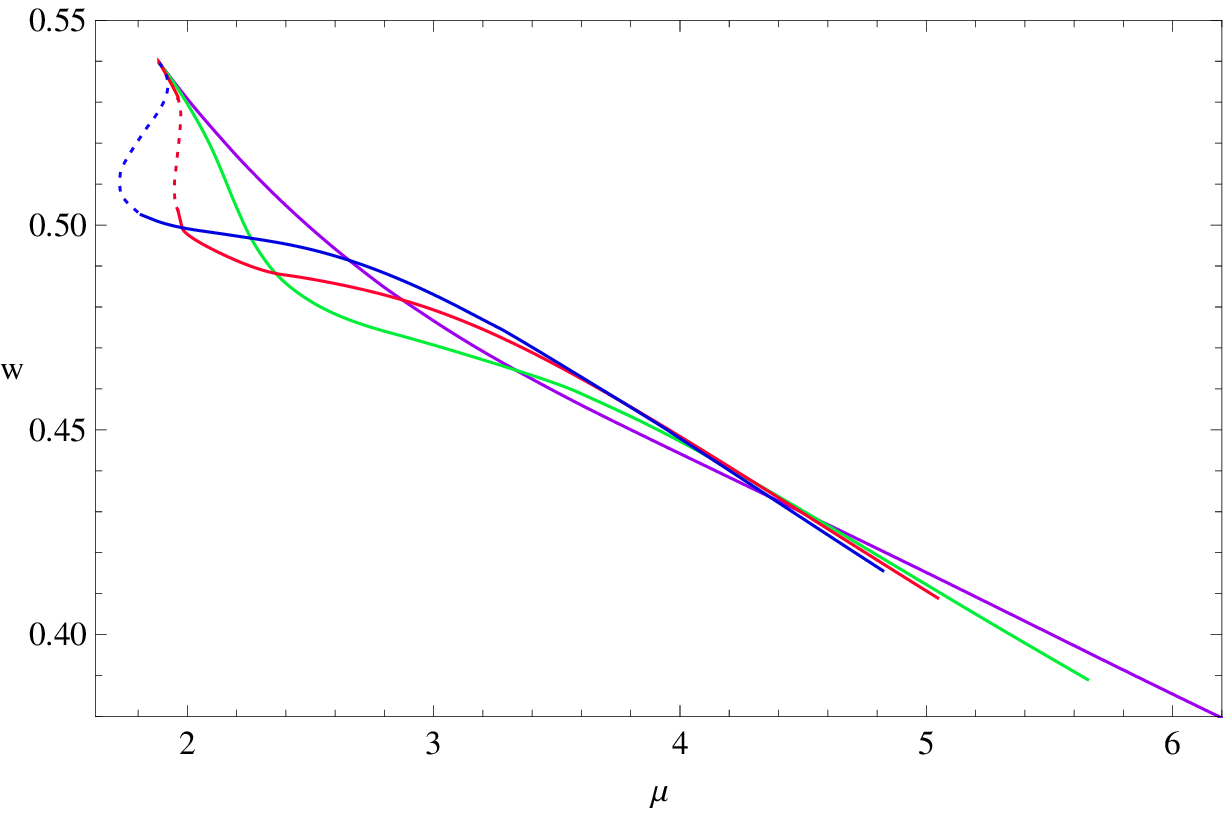}
\caption{\label{xylittle} The pseudo potential as a function of
chemical potential at fixed $\ell/L=0.9$ (left plot) and $\ell/L=2$
(right plot). Lines  are for $\zeta=0$ (purple), $\zeta=0.2$
(green), $\zeta=0.65$ (red) and $\zeta=1.6$ (blue), respectively.
The dotted part of the lines for $\zeta=0.65$ (red) and $\zeta=1.6$
(blue) are not thermodynamically favored, trace the
 physical curves by always choosing the solid lines.}
\end{figure}

But we find that the behavior of the  pseudo potential with the
change of chemical potential is different from the case in the
temporal Wilson loop. In Figure~\eqref{xylittle} we plot the pseudo
potentials with respect to chemical potential for fixed distance
$\ell/L=0.9$ (in the deconfinement phase) and $\ell/L=2$ (in the
confinement phase). In the case of $\ell/L=0.9$, we see that as
$\zeta$ is not very large, the behavior of heavy quark potential  is
qualitatively similar to each other, no matter the order of the
phase transition. At the beginning of the phase transition, it
firstly decrease in the superconducting phase and then increases
continuously for large chemical potential. The position of the
bottom moves to small chemical potential side with the increase of
$\zeta$. For sufficient large $\zeta$, the bottom is cut off and the
 pseudo potential increases monotonically with the
increase of the chemical potential.

%
%
%
%

\section{Conclusion and Discussions}
\label{sect:conclusion}

To better understand the properties of holographic superconductor,
we investigated the behaviors of two non-local physical quantities,
i.e., entanglement entropy and Wilson loop in the St\"{u}ckelberg
holographic insulator/superconductor model. The model exhibits rich
phase structure. For small $\zeta$, the transition at $\mu_c$ is
second order and no discontinuity within the superconducting phase.
For intermediate $\zeta$, there is an additional complication. As we
can see from the left plot of Figure~\eqref{smfree}, the grand
potential as a function of chemical potential develops a ``swallow
tail" at a certain chemical potential $\mu_0$ in the superconducting
phase. Thus, there is a jump in the condensate within the
superconducting phase. For large $\zeta$, the superconducting phase
transition becomes first order.

We calculated the entanglement entropy for a strip geometry and the
VEV for temporal Wilson loop and spatial Wilson loop in the
insulator/superconductor phase transition. Both show that there
exists a confinement/deconfinement phase transition (see
Figure~\eqref{entropywidth} and Figure~\eqref{xtlength}). In both
cases, we found that the critical width for the strip and the
critical distance between quark and antiquark are non-monotonic
functions of chemical potential (see Figure~\eqref{heelc} and
Figure~\eqref{xtlc}).

There is either a discontinuity of the entanglement entropy or its
slop with respect to chemical potential at $\mu_c$, namely, at the
insulator/supercondcutor transition point, which indicates some kind
of significant reorganization of the degrees of freedom as new
degrees of freedom are expected to emerge in the superconducting
phase. These discontinuities of the entanglement entropy or its slop
correctly indicate the order of associated phase transition. Beyond
the critical chemical potential, we found that the entanglement
entropy is not monotonic in the
  superconducting phase: at the beginning of the transition, the entropy increases and reaches its maximum
   at a certain chemical potential $\mu_{max}$ and then decreases monotonically. This behavior is universal in
   the holographic insulator/superconductor model. As one increases the model parameter
     $\zeta$, but keeps the parameter $\beta$, the position of the peak $\mu_{max}$ moves to the small chemical
     potential side.

  Compared to the phenomenon observed in the entanglement
 entropy, the (pseudo) heavy quark potential with respect to chemical potential in superconducting phase
  presents more rich behaviors. For a fixed chemical potential, the
  heavy quark potential always increases with the
  distance between the quark-antiquark pair (see
  Figure~\eqref{xtlength}). In the small $\ell$ region the potential
  has a Coulomb potential form, while in the large $\ell$ region,
  the potential shows a confining potential form, as it is expected.
  But the behaviors of the heavy quark potential with respect to
  chemical potential are just opposite to each other in the small
  $\ell$ region and in the large $\ell$ region.

In the spatial Wilson loop case, the pseudo potential shows a
non-monotonic behavior with respect to chemical potential in the
small $\ell$ region, but not in the large $\ell$ region (see
Figure~\eqref{xylittle}). The non-monotonic behavior might be
related to the one in the entanglement entropy.

In this paper, we just reported the behaviors of the entanglement
entropy and heavy quark potential in the holographic St\"{u}ckelberg
insulator/superconductor model. Unfortunately, due to
the lack of the knowledge of the dual field theory in the ``bottom-up" approach, we can not
completely understand some implications of the behaviors of entanglement entropy and heavy quark
potential, in particular, for the non-monotonic behavior, although we tried to give some discussions
for this in Ref.~\cite{Cai:2012sk}. But anyway, different from
some physical quantities, such as condensate and charge density, in superconducting phase,
we can see that there are various phase structures in superconducting phase by investigating
the behaviors of entanglement entropy and Wilson loop.  It turns out that these two non-local physical
quantities are good probes to study the properties of the
holographic phase transition.

In this paper we limited ourselves to the case with $\mathcal{F}(\psi)=\psi^2+\zeta\psi^6$. For other form of the
function, see, for example, Refs.\cite{Franco:2009yz,Aprile:2009ai,Peng:2011gh}, we expect that
the result will be qualitatively same. But it is interesting to
confirm this. In addition,  It would be of great interest to
investigate the behavior of entanglement entropy in the holographic
p-wave insulator/superconductor case, in order to see whether
the non-monotonic behavior is more universal or not.

\section*{Acknowledgements}
We thank T. Takayanagi for helpful correspondences and in particular
for his suggestion to calculate the heavy quark potential in the
holographic insulator/superconductor phase transition. This work was
supported in part by the National Natural Science Foundation of
China (No.10821504, No.10975168, No.11035008 and No.11205226), and
in part by the Ministry of Science and Technology of China under
Grant No. 2010CB833004. SH and LFL would like appreciate the general
financial support from China Postdoctoral Science Foundation No.
2012M510562 and No. 2012M510563 respectively. SH also would like
thank for Department of Electrophysics of National Chiao-Tung
University, National Taiwan university and the "International School
On Strings And Fundamental Physics" held at Hamburg, for their
hospitality and financial support. The authors are grateful to
Zhang-Yu Nie, Wen-Yu Wen, Yi Yang, Hai-Qing Zhang and Yun-Long Zhang
for useful discussions.

\appendix


\begin{thebibliography}{99}



\baselineskip 12pt

\bibitem{2006PhRvB..73x5115R}
Ryu, S.,  Hatsugai, Y.
``Entanglement entropy and the Berry phase in the solid state,"  Phys.\ Rev.\  B {\bf 73}, 245115 (2006)
[arXiv:cond-mat/0601237].

\bibitem{Amico:2007ag}
  L.~Amico, R.~Fazio, A.~Osterloh and V.~Vedral,
  ``Entanglement in many-body systems,''  Rev.\ Mod.\ Phys.\  {\bf 80}, 517 (2008)  [quant-ph/0703044 [QUANT-PH]].


\bibitem{Maldacena:1997re}
  J.~M.~Maldacena,
  ``The large N limit of superconformal field theories and supergravity,''
  Adv.\ Theor.\ Math.\ Phys.\  {\bf 2}, 231 (1998)
  [Int.\ J.\ Theor.\ Phys.\  {\bf 38}, 1113 (1999)]
  [arXiv:hep-th/9711200].
\bibitem{Gubser:1998bc}
  S.~S.~Gubser, I.~R.~Klebanov and A.~M.~Polyakov,
  ``Gauge theory correlators from non-critical string theory,''
  Phys.\ Lett.\  B {\bf 428}, 105 (1998)
  [arXiv:hep-th/9802109].
\bibitem{Witten:1998qj}
  E.~Witten,
  ``Anti-de Sitter space and holography,''
  Adv.\ Theor.\ Math.\ Phys.\  {\bf 2}, 253 (1998)
  [arXiv:hep-th/9802150].

\bibitem{Aharony:1999ti}
  O.~Aharony, S.~S.~Gubser, J.~M.~Maldacena, H.~Ooguri and Y.~Oz,
  Phys.\ Rept.\  {\bf 323}, 183 (2000)
  [arXiv:hep-th/9905111].


\bibitem{Ryu:2006bv}
  S.~Ryu and T.~Takayanagi,
  ``Holographic derivation of entanglement entropy from AdS/CFT,''  Phys.\ Rev.\ Lett.\  {\bf 96}, 181602 (2006)  [hep-th/0603001].

\bibitem{Nishioka:2009un}
  T.~Nishioka, S.~Ryu and T.~Takayanagi,
  ``Holographic Entanglement Entropy: An Overview,''  J.\ Phys.\ A A {\bf 42}, 504008 (2009)  [arXiv:0905.0932 [hep-th]].

\bibitem{Takayanagi:2012kg}
  T.~Takayanagi,
  ``Entanglement Entropy from a Holographic Viewpoint,''  arXiv:1204.2450 [gr-qc].


\bibitem{Hartnoll:2008vx}
  S.~A.~Hartnoll, C.~P.~Herzog and G.~T.~Horowitz,
  ``Building a Holographic Superconductor,''
  Phys.\ Rev.\ Lett.\  {\bf 101}, 031601 (2008)
  [arXiv:0803.3295 [hep-th]].

\bibitem{Albash:2008eh}
  T.~Albash and C.~V.~Johnson,
  ``A Holographic Superconductor in an External Magnetic Field,''  JHEP {\bf 0809}, 121 (2008)  [arXiv:0804.3466 [hep-th]].


\bibitem{Cai:2011tm}
  R.~-G.~Cai, L.~Li, H.~-Q.~Zhang and Y.~-L.~Zhang,
  ``Magnetic Field Effect on the Phase Transition in AdS Soliton Spacetime,''  Phys.\ Rev.\ D {\bf 84}, 126008 (2011)  [arXiv:1109.5885 [hep-th]].  

\bibitem{Horowitz:2008bn}
  G.~T.~Horowitz and M.~M.~Roberts,
  ``Holographic Superconductors with Various Condensates,''  Phys.\ Rev.\ D {\bf 78}, 126008 (2008)  [arXiv:0810.1077 [hep-th]].

\bibitem{Brynjolfsson:2009ct}
  E.~J.~Brynjolfsson, U.~H.~Danielsson, L.~Thorlacius and T.~Zingg,
  ``Holographic Superconductors with Lifshitz Scaling,''  J.\ Phys.\ A A {\bf 43}, 065401 (2010)  [arXiv:0908.2611 [hep-th]].

\bibitem{Cai:2009hn}
  R.~-G.~Cai and H.~-Q.~Zhang,
  ``Holographic Superconductors with Horava-Lifshitz Black Holes,''  Phys.\ Rev.\ D {\bf 81}, 066003 (2010)  [arXiv:0911.4867 [hep-th]].

\bibitem{Horowitz:2009ij}
  G.~T.~Horowitz and M.~M.~Roberts,
  ``Zero Temperature Limit of Holographic Superconductors,''  JHEP {\bf 0911}, 015 (2009)  [arXiv:0908.3677 [hep-th]].

\bibitem{Horowitz:2011dz}
  G.~T.~Horowitz, J.~E.~Santos and B.~Way,
  ``A Holographic Josephson Junction,''  Phys.\ Rev.\ Lett.\  {\bf 106}, 221601 (2011)  [arXiv:1101.3326 [hep-th]].

\bibitem{Wang:2012yj}
  Y.~-Q.~Wang, Y.~-X.~Liu, R.~-G.~Cai, S.~Takeuchi and H.~-Q.~Zhang,
``Holographic SIS Josephson Junction,''  arXiv:1205.4406 [hep-th].


\bibitem{Montull:2011im}
  M.~Montull, O.~Pujolas, A.~Salvio and P.~J.~Silva,
  ``Flux Periodicities and Quantum Hair on Holographic Superconductors,''  Phys.\ Rev.\ Lett.\  {\bf 107}, 181601 (2011)  [arXiv:1105.5392 [hep-th]].  


\bibitem{Bobev:2011rv}
  N.~Bobev, A.~Kundu, K.~Pilch and N.~P.~Warner,
  ``Minimal Holographic Superconductors from Maximal Supergravity,''  JHEP {\bf 1203}, 064 (2012)  [arXiv:1110.3454 [hep-th]].


\bibitem{Liu:2012hc}
  Y.~Liu, Y.~Peng and B.~Wang,
  ``Gauss-Bonnet holographic superconductors in Born-Infeld electrodynamics with backreactions,''  arXiv:1202.3586 [hep-th].

\bibitem{Erdmenger:2011tj}
  J.~Erdmenger, P.~Kerner and H.~Zeller,
  ``Transport in Anisotropic Superfluids: A Holographic Description,''  JHEP {\bf 1201}, 059 (2012)  [arXiv:1110.0007 [hep-th]].


\bibitem{Gubser:2008wv}
S.~S.~Gubser and S.~S.~Pufu,
``The Gravity dual of a p-wave superconductor,''  JHEP {\bf 0811}, 033 (2008)  [arXiv:0805.2960 [hep-th]].

\bibitem{Hartnoll:2012pp}
  S.~A.~Hartnoll and R.~Pourhasan,
``Entropy balance in holographic superconductors,''  JHEP {\bf 1207}, 114 (2012)  [arXiv:1205.1536 [hep-th]].

\bibitem{Erdmenger:2012ik}
  J.~Erdmenger, P.~Kerner and S.~Muller,
``Towards a Holographic Realization of Homes' Law,''  arXiv:1206.5305 [hep-th].

\bibitem{Montull:2012fy}
  M.~Montull, O.~Pujolas, A.~Salvio and P.~J.~Silva,
``Magnetic Response in the Holographic Insulator/Superconductor Transition,''  JHEP {\bf 1204}, 135 (2012)  [arXiv:1202.0006 [hep-th]].

\bibitem{Barclay:2010up}
  L.~Barclay, R.~Gregory, S.~Kanno and P.~Sutcliffe,
``Gauss-Bonnet Holographic Superconductors,''  JHEP {\bf 1012}, 029
(2010)  [arXiv:1009.1991 [hep-th]].

\bibitem{Cai:2010cv}
  R.~-G.~Cai, Z.~-Y.~Nie and H.~-Q.~Zhang,
  ``Holographic p-wave superconductors from Gauss-Bonnet gravity,''
  Phys.\ Rev.\ D {\bf 82}, 066007 (2010)
  [arXiv:1007.3321 [hep-th]].

\bibitem{Siani:2010uw}
  M.~Siani,
``Holographic Superconductors and Higher Curvature Corrections,''
JHEP {\bf 1012}, 035 (2010)  [arXiv:1010.0700 [hep-th]].

\bibitem{Cai:2010zm}
  R.~-G.~Cai, Z.~-Y.~Nie and H.~-Q.~Zhang,
  ``Holographic Phase Transitions of P-wave Superconductors in Gauss-Bonnet Gravity with Back-reaction,''  Phys.\ Rev.\ D {\bf 83}, 066013 (2011)  [arXiv:1012.5559 [hep-th]].

\bibitem{Wu:2010vr}
  J.~-P.~Wu, Y.~Cao, X.~-M.~Kuang and W.~-J.~Li,
``The 3+1 holographic superconductor with Weyl corrections,''
Phys.\ Lett.\ B {\bf 697}, 153 (2011)  [arXiv:1010.1929 [hep-th]].



\bibitem{Albash:2012pd}
T.~Albash and C.~V.~Johnson,
``Holographic Studies of Entanglement Entropy in Superconductors,''  JHEP {\bf 1205}, 079 (2012)  [arXiv:1202.2605 [hep-th]].

\bibitem{Cai:2012nm}
R.~-G.~Cai, S.~He, L.~Li and Y.~-L.~Zhang,
``Holographic Entanglement Entropy on P-wave Superconductor Phase Transition,''  JHEP {\bf 1207}, 027 (2012)  [arXiv:1204.5962 [hep-th]].


\bibitem{Cai:2012sk}
R.~-G.~Cai, S.~He, L.~Li and Y.~-L.~Zhang,
``Holographic Entanglement Entropy in Insulator/Superconductor Transition,''  JHEP {\bf 1207}, 088 (2012)  [arXiv:1203.6620 [hep-th]].












\bibitem{Nishioka:2009zj}
  T.~Nishioka, S.~Ryu and T.~Takayanagi,
  ``Holographic Superconductor/Insulator Transition at Zero Temperature,''  JHEP {\bf 1003}, 131 (2010)  [arXiv:0911.0962 [hep-th]].





\bibitem{Franco:2009yz}
  S.~Franco, A.~Garcia-Garcia and D.~Rodriguez-Gomez,
``A General class of holographic superconductors,''  JHEP {\bf 1004}, 092 (2010)  [arXiv:0906.1214 [hep-th]].


\bibitem{Horowitz:2010jq}
  G.~T.~Horowitz and B.~Way,
``Complete Phase Diagrams for a Holographic Superconductor/Insulator
System,''  JHEP {\bf 1011}, 011 (2010)  [arXiv:1007.3714 [hep-th]].

\bibitem{Aprile:2009ai}
  F.~Aprile and J.~G.~Russo,
``Models of Holographic superconductivity,''  Phys.\ Rev.\ D {\bf 81}, 026009 (2010)  [arXiv:0912.0480 [hep-th]].


\bibitem{Peng:2011gh}
  Y.~Peng, Q.~Pan and B.~Wang,
``Various types of phase transitions in the AdS soliton background,''  Phys.\ Lett.\ B {\bf 699}, 383 (2011)  [arXiv:1104.2478 [hep-th]].


\bibitem{con1}  T.~Nishioka and T.~Takayanagi,
  ``AdS Bubbles, Entropy and Closed String Tachyons,''
  JHEP {\bf 0701}, 090 (2007)
  [hep-th/0611035].

\bibitem{con2}I.~R.~Klebanov, D.~Kutasov and A.~Murugan,
  ``Entanglement as a probe of confinement,''
  Nucl.\ Phys.\ B {\bf 796}, 274 (2008)
  [arXiv:0709.2140 [hep-th]].

\bibitem{Myers:2012ed}
  R.~C.~Myers and A.~Singh,
``Comments on Holographic Entanglement Entropy and RG Flows,''  JHEP {\bf 1204}, 122 (2012)  [arXiv:1202.2068 [hep-th]].






\bibitem{Maldacena:1998im}
  J.~M.~Maldacena,
  ``Wilson loops in large N field theories,''
  Phys.\ Rev.\ Lett.\  {\bf 80}, 4859 (1998)
  [arXiv:hep-th/9803002].

\bibitem{Rey:1998bq}
  S.~J.~Rey, S.~Theisen and J.~T.~Yee,
  ``Wilson-Polyakov loop at finite temperature in large N gauge theory and
  anti-de Sitter supergravity,''
  Nucl.\ Phys.\  B {\bf 527}, 171 (1998)
  [arXiv:hep-th/9803135].


\bibitem{Polyakov:1997tj}
  A.~M.~Polyakov,
  ``String theory and quark confinement,''
  Nucl.\ Phys.\ Proc.\ Suppl.\  {\bf 68}, 1 (1998)
  [arXiv:hep-th/9711002].

\bibitem{Cai:2012xh}
  R.~-G.~Cai, S.~He and D.~Li,
  ``A hQCD model and its phase diagram in Einstein-Maxwell-Dilaton system,''  JHEP {\bf 1203}, 033 (2012)  [arXiv:1201.0820 [hep-th]].  

\bibitem{Andreev-T1}
  O.~Andreev and V.~I.~Zakharov,
  ``The Spatial String Tension, Thermal Phase Transition, and AdS/QCD,''
  Phys.\ Lett.\  B {\bf 645}, 437 (2007)
  [arXiv:hep-ph/0607026];
  O.~Andreev,
  ``The Spatial String Tension in the Deconfined Phase of SU(N) Gauge Theory
  and Gauge/String Duality,''
  Phys.\ Lett.\  B {\bf 659}, 416 (2008)
  [arXiv:0709.4395 [hep-ph]].

\bibitem{Li:2011hp}
  D.~Li, S.~He, M.~Huang and Q.~-S.~Yan,
  ``Thermodynamics of deformed AdS$_5$ model with a positive/negative quadratic correction in graviton-dilaton system,''  JHEP {\bf 1109}, 041 (2011)  [arXiv:1103.5389 [hep-th]].  






\end{thebibliography}
\end{document}